\documentclass[sigconf,natbib=true]{acmart}
\AtBeginDocument{%
  }

\usepackage{hyperref}       
\usepackage{amsfonts}       
\usepackage{bm}
\usepackage{graphicx}
\usepackage{amsthm}
\usepackage{amsmath}
\usepackage{subfigure}
\usepackage{multirow}
\usepackage{enumitem}
\usepackage{bbding}
\usepackage{colortbl}
\usepackage{arydshln}
\usepackage{algorithm}  
\usepackage{algorithmicx}  
\usepackage{threeparttable}
\usepackage{algpseudocode}  
\usepackage{graphicx}
\usepackage{subcaption}

\setcopyright{acmlicensed}
\copyrightyear{2018}
\acmYear{2018}
\acmDOI{XXXXXXX.XXXXXXX}
\acmConference[Conference acronym 'XX]{Make sure to enter the correct
  conference title from your rights confirmation emai}{June 03--05,
  2018}{Woodstock, NY}
\acmISBN{978-1-4503-XXXX-X/18/06}

\begin{document}

\title{QE-RAG: A Robust Retrieval-Augmented Generation Benchmark \mbox{for Query Entry Errors}}

\author{Kepu Zhang}
\author{Zhongxiang Sun}
\affiliation{
  \institution{\mbox{Gaoling School of Artificial Intelligence}\\Renmin University of China} \city{Beijing}\country{China}
  } 
\email{kepuzhang@ruc.edu.cn}

\author{Weijie Yu}
\affiliation{
  \institution{School of Information Technology 
and Management\\University of International Business
and Economics} \city{Beijing}\country{China}
  }

\author{Xiaoxue Zang}
\author{Kai Zheng}
\affiliation{
  \institution{Kuaishou Technology Co., Ltd.}
  \city{Beijing}\country{China}
  }

\author{Yang Song}
\author{Han Li}
\affiliation{
  \institution{Kuaishou Technology Co., Ltd.}
  \city{Beijing}\country{China}
  }

\author{Jun Xu}
\affiliation{
  \institution{\mbox{Gaoling School of Artificial Intelligence}\\Renmin University of China} \city{Beijing}\country{China}
  }

\renewcommand{\shortauthors}{Kepu Zhang et al.}

\begin{abstract}
Retriever-augmented generation (RAG) has become a widely adopted approach for enhancing the factual accuracy of large language models (LLMs).
While current benchmarks evaluate the performance of RAG methods from various perspectives, they share a common assumption that user queries used for retrieval are error-free. However, in real-world interactions between users and LLMs, query entry errors such as keyboard proximity errors, visual similarity errors, and spelling errors are frequent. The impact of these errors on current RAG methods against such errors remains largely unexplored. 
To bridge this gap, we propose QE-RAG, the first robust RAG benchmark designed specifically to evaluate performance against query entry errors. 
We augment six widely used datasets by injecting three common types of query entry errors into randomly selected user queries at rates of 20\% and 40\%, simulating typical user behavior in real-world scenarios.
We analyze the impact of these errors on LLM outputs and find that corrupted queries degrade model performance, which can be mitigated through query correction and training a robust retriever for retrieving relevant documents.
Based on these insights, we propose a contrastive learning-based robust retriever training method and a retrieval-augmented query correction method.
Extensive in-domain and cross-domain experiments reveal that:
(1) state-of-the-art RAG methods including sequential, branching, and iterative methods, exhibit poor robustness to query entry errors;
(2) our method significantly enhances the robustness of RAG when handling query entry errors and it's compatible with existing RAG methods, further improving their robustness.
\end{abstract}




\maketitle

\section{Introduction}\label{sec:intro}
Retriever-augmented generation (RAG)~\cite{borgeaud2022improving,lewis2020retrieval,chen2024benchmarking}, which integrates retrieval mechanisms to incorporate external knowledge into large language models (LLMs), has become a widely adopted approach. By retrieving knowledge from external sources, RAG addresses issues such as insufficient knowledge and hallucinations in LLMs~\cite{tonmoy2024comprehensive,gao2023retrieval}, thereby improving the accuracy and fidelity of their responses.
\begin{figure}[t]
    \centering
\includegraphics[width=\linewidth]{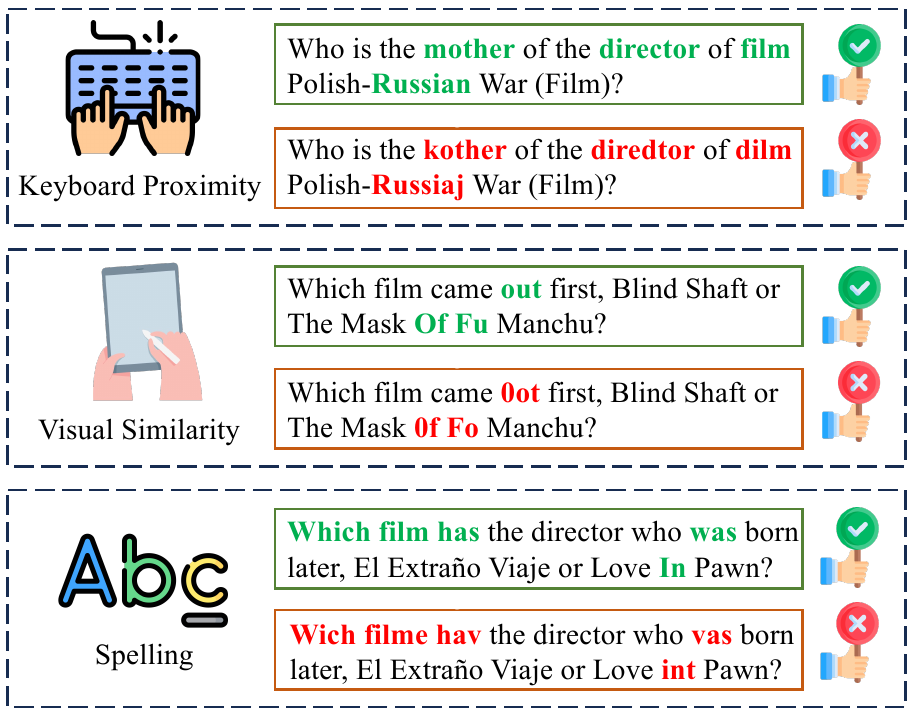  }
    \caption{Examples of three types of query entry errors including keyboard proximity errors, visual similarity
errors, and spelling errors.
    }
    \label{fig:fig1}
\end{figure}

Current RAG benchmarks evaluate the performance of RAG methods from various perspectives. For example,~\citet{es2024ragas} assess fidelity in LLM-generated content,~\citet{chen2024benchmarking} evaluate the model’s ability to refuse to answer inappropriate or unanswerable queries, and~\citet{liu2023recall} examine the capacity of models to handle counterfactual information. Although these studies provide valuable insights into model effectiveness across different scenarios, they universally assume that user queries are error-free. In real-world settings, as illustrated in Figure~\ref{fig:fig1}, user queries often contain entry errors such as keyboard proximity errors, visual similarity errors, and spelling mistakes. The impact of these errors on LLM outputs remains largely unexplored.

To fill this gap, we introduce QE-RAG, the first RAG benchmark specifically designed to evaluate model performance under query entry errors. We inject three common types of query errors—spelling errors, keyboard proximity errors, and visual similarity errors—into four direct QA datasets (TriviaQA~\cite{joshi2017triviaqa}, Natural Questions~\cite{kwiatkowski2019natural}, PopQA~\cite{mallen2022not}, and WebQuestions~\cite{berant2013semantic}) and two multi-hop QA datasets (HotpotQA~\cite{yang2018hotpotqa} and 2WikiMultiHopQA~\cite{ho2020constructing}). Specifically, 
we use the nlpaug~\cite{ma2019nlpaug} tool to systematically inject these errors, applying them in a 3:1:1 ratio to reflect real-world error distribution patterns. For each query, there is a 30\% probability of selecting a word, and for each selected word, a 30\% probability of corrupting a character. This setup realistically simulates typical user query behaviors, providing a practical evaluation environment for RAG models. Since these errors do not alter the user’s underlying information need, we retain the original RAG labels for the corrupted queries. To simulate varying levels of noise, we generate two versions of the QE-RAG by corrupting 20\% and 40\% of the queries, representing moderate and high-error scenarios.

Based on the proposed QE-RAG dataset, we conducted preliminary experiments (\textsection~\ref{sec:pre experiments}) on the corrupted HotpotQA and Natural Questions (NQ) datasets to explore the impact of query entry errors on LLM outputs. 
We find that: (1) \textbf{Retrieving correct documents} for corrupted queries can enhance the RAG model's robustness to query entry errors. (2) \textbf{Correcting corrupted queries} also improves the RAG model's robustness.
Therefore, (1) To retrieve correct documents, we train a robust retriever using contrastive learning based on a retrieval dataset with a 20\% error query rate, enabling it to retrieve the correct document corresponding to the correct query even when faced with corrupted queries. (2) To correct corrupted queries, we adopt the current state-of-the-art LLM-based correction methods. However, considering the significant issue of overcorrection~\cite{li2023effectiveness,fang2023chatgpt} in LLMs during correction and LLMs may have limitations in recognizing certain uncommon knowledge during query correction~\cite{zhang2024trigger}, we propose a query correction approach that combines RAG (based on the robust retriever we introduced earlier) with fine-tuning to mitigate overcorrection while enhancing robustness.

We selected the state-of-the-art retriever BGE \cite{xiao2023c} from the MTEB leaderboard \cite{muennighoff2022mteb} and two large language models, Qwen2 \cite{yang2024qwen2} and LLama3 \cite{llama3modelcard}, to evaluate their robustness to query entry errors.
We tested the in-domain and cross-domain performance of various existing RAG methods (e.g., trained on HotpotQA and tested on the same or other datasets) to assess their robustness against query entry errors. 
These RAG methods include standard RAG~\cite{gao2023retrieval}, query reformulation~\cite{gao2023precise}, document refinement~\cite{jiang2023longllmlingua}, branching~\cite{shi2024replug,kimsure} and iterative~\cite{shao2023enhancing} methods.

Extensive experimental results show that while these state-of-the-art RAG methods demonstrate some effectiveness compared to standard RAG, their robustness to query entry errors remains limited. In contrast, the two methods we propose significantly enhance the robustness of RAG systems and can be combined with existing RAG methods to further improve their performance.

To summarize, our contributions are as follows: 
\begin{itemize} 
\item To the best of our knowledge, we are the first to investigate robustness against query entry errors in RAG research, focusing on three representative error types: keyboard proximity, visual similarity, and spelling. 
\item We construct a benchmark dataset, QE-RAG, based on six widely-used RAG datasets, incorporating two levels of noise through the explicit injection of three types of errors. Extensive experiments conducted on QE-RAG demonstrate that state-of-the-art RAG methods, including query reformulation, document refinement, branching, and iterative methods, exhibit poor robustness to query entry errors.
\item We propose two solutions to improve robustness against query entry errors: (1) a contrastive learning-based trained robust retriever, which enhances RAG robustness; (2) a retrieval-augmented query correction method, resulting in further improvements in robustness. 
\end{itemize}

\section{Related Work}\label{sec:intro}
\begin{table*}[t]
\centering
\caption{The statistics of six datasets used in QE-RAG. ``Source'' refers to the knowledge source of each dataset. ``\#Query'' denotes the number of queries. ``0\% Prob'', ``20\% Prob'', ``40\% Prob'' represent the proportions of corrupted queries in the dataset at 0\%, 20\%, and 40\%, respectively. ``Avg. \#Char/Query'' indicates the average number of characters per query. ``Avg. \#Words/Query'' refers to the average number of words per query.}
\resizebox{0.9\textwidth}{!}{
\begin{tabular}{llcc|ccc| ccc}
    \toprule
    \multirow{2}{*}{Type} &\multirow{2}{*}{Dataset} &\multirow{2}{*}{Source} &\multirow{2}{*}{\#Query}  &\multicolumn{3}{c}{\multirow{1}{*}{Avg. \#Chars/Query}} &\multicolumn{3}{c}{\multirow{1}{*}{Avg. \#Words/Query}}\\
    \cmidrule{5-10} 
    & &  & &0\% Prob &20\% Prob &40\% Prob &0\% Prob &20\% Prob &40\% Prob \\ 
    \midrule
    \multirow{4}{*}{QA}
    &NQ&Wiki&3610&48.4&48.6&48.7&9.4&9.4&9.4\\
    &PopQA&Wiki&14267&37.1&37.4&37.7&6.7&6.8&6.9\\
    &TrivalQA&Wiki \& Web&11313&69.1&69.4&69.6&12.6&12.6&12.7\\
    &WebQA&Google Freebase&2032&38.0&38.0&38.1&6.8&6.9&6.9\\
    \hline
    \multirow{2}{*}{Multi-Hop QA}
    &HotpotQA&Wiki&7405 &94.5&94.8&95.1&16.4&16.4&16.5\\
    &2wiki&Wiki&12576&68.1&68.5&68.8&12.4&12.5&12.5 \\
    \bottomrule
\end{tabular}
}

\label{tab:data_main}
\end{table*}

\subsection{RAG benchmark}
QA datasets can be widely used to test the effects of RAG models, promoting the development of RAG technology. 
Among them, the Natural Questions (NQ)~\cite{kwiatkowski2019natural} is an open-domain question-answering dataset. HotpotQA~\cite{yang2018hotpotqa} and 2WikiMultiHopQA~\cite{ho2020constructing} are multi-hop reasoning datasets, requiring stronger reasoning abilities of the model. The TriviaQA~\cite{joshi2017triviaqa} dataset has a large syntactic and lexical difference between the question and the answer. The PopQA~\cite{mallen2022not} dataset supplements the long-tail information that may be missed in the process QA dataset, etc. 
Above datasets 
have been manually annotated and reviewed, without incorporating query entry errors into the dataset.
Existing RAG benchmarks primarily assess the quality of content generated by LLMs or the LLM's ability to process external information. RAGAS~\cite{es2024ragas} and ARES~\cite{saad2024ares} evaluate the contextual relevance and fidelity of LLM-generated content. RGB~\cite{chen2024benchmarking} tests the robustness of LLM against noisy documents and the ability to refuse to answer, while RECALL~\cite{liu2023recall} analyzes the LLM's processing capability regarding counterfactual information. 
However, they all assume that the queries used for retrieval are correct, without considering the actual scenarios where users may enter corrupted queries.
In the increasingly popular era of LLM, this cannot well evaluate the real capabilities of RAG technology. Therefore, this paper focuses on establishing an RAG evaluation framework that includes corrupted queries, which can help evaluate the robustness of RAG models and promote the further development of RAG technology in the era of LLM.

\subsection{Retriever Augmented Generation }
In the era of Language Models (LLMs), the way of obtaining information has changed, with more and more users preferring to obtain information through LLM rather than search engines~\cite{ai2023information,dai2024neural}. 
By combining information retrieval and generation, the emergence of RAG technology allows LLM to gain new knowledge from external databases as a supplement, making its generated content more accurate and reliable~\cite{lewis2020retrieval,gao2023retrieval}.
Standard RAG methods~\cite{gao2023retrieval} supplement user queries with retrieved documents, which are then fed into the LLM to generate responses. Over time, numerous approaches have been proposed to further enhance the performance of RAG systems. Following~\cite{jin2024flashrag}, these methods can be categorized into sequential pipeline, branching pipeline, iterative pipeline, and so on.

In the sequential pipeline, query reformulation methods focus on improving the input query to optimize the retrieval process. These techniques operate under the assumption that user queries may not always be optimal for retrieval tasks: HyDE~\cite{gao2023precise}: The LLM generates a hypothetical document based on the query, which is then used as the query for retrieval. This approach assumes that the generated document aligns better with the retrieved documents.
Query2doc~\cite{wang2023query2doc} concatenates the LLM-generated pseudo-document with the original query to form a new query for retrieval. Rewrite-Retrieve-Read~\cite{ma2023query} proposes fine-tuning a query rewriter to optimize query reformulation. BEQUE~\cite{ye2023improving} employs a combination of fine-tuning and reinforcement learning to rewrite queries, particularly improving retrieval performance for long-tail queries. The above query reformulation methods do not consider that the query itself is corrupted, thus ignoring that query reformulation may accumulate and amplify errors, which will seriously affect the final RAG performance.
Another line of work involves processing the retrieved documents to make them more useful for the LLM: Selective-Content~\cite{li2023compressing} compresses the provided context by removing redundant information using self-information metrics.
LLMLingua~\cite{jiang2023llmlingua} uses smaller models to detect and remove unnecessary tokens in the prompt, making the remaining content more interpretable for the LLM (even if humans may find it less comprehensible).
LongLLMLingua~\cite{jiang2023longllmlingua} extends LLMLingua by incorporating question-aware techniques to extract key information from retrieved documents, improving their alignment with the LLM's processing capabilities.

Branching pipelines process multiple paths in parallel to enhance performance: 
REPLUG~\cite{shi2024replug} integrates document relevance into the LLM's response generation, improving the accuracy and contextual alignment of generated outputs.
SuRe~\cite{kimsure} utilizes summarization techniques to select the most suitable answer from multiple candidate responses.
Iterative pipelines aim to refine the retrieval process dynamically
Iter-RetGen~\cite{shao2023enhancing} enhances the retrieval query by iteratively incorporating LLM responses into the query, leveraging the generated feedback to refine retrieval results.
In this paper, we will evaluate the robustness of these state-of-the-art RAG methods in scenarios where queries contain errors.

\section{QE-RAG Dataset Construction}\label{sec:main sec}
\begin{figure*}
    \centering
\includegraphics[width=0.98\linewidth]{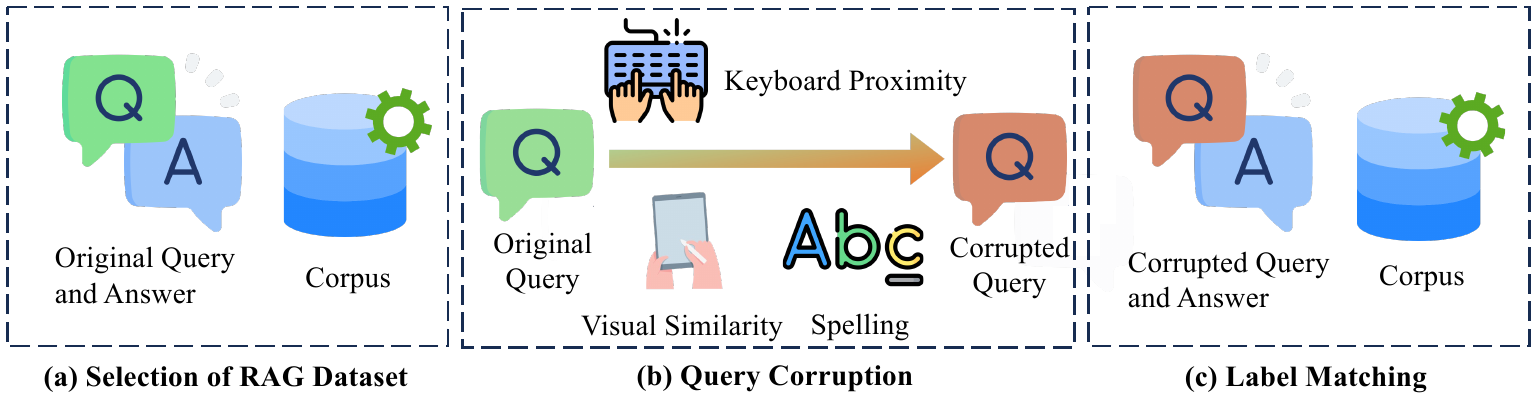}
    \caption{
    The construction process of QE-RAG datasets. (a) Selection of RAG Dataset. (b) Query corruption through three scenarios: keyboard proximity errors, visual similarity errors and spelling errors. (c) Label matching.
    }
    \label{fig:data construct}
\end{figure*}

We focus on RAG in this study, which is formulated as follows: given a query \( q \in \mathcal{Q} \) (where \( \mathcal{Q} \) is the set of all possible queries) and an external knowledge base \( K = \{d_1, d_2, \dots, d_N\} \) consisting of \( N \) documents, the goal of RAG is to generate a response \( a \in \mathcal{A} \) (where \( \mathcal{A} \) is the set of possible answers) by leveraging both retrieval from the knowledge base and generation from a LLM. 
 Unlike previous datasets, which assume that 
$q$ is error-free, we consider a more practical scenario in which $q$ may be corrupted by three types of query entry errors. As illustrated in Figure~\ref{fig:data construct}, our QE-RAG dataset is constructed through the following steps.

\textbf{Step1: Selection of RAG Dataset.}
Following FlashRAG~\cite{jin2024flashrag}, we collect and extend six widely-used RAG datasets to form our \textbf{QE-RAG}, which includes four direct QA datasets (TriviaQA~\cite{joshi2017triviaqa}, Natural Questions~\cite{kwiatkowski2019natural}, PopQA~\cite{mallen2022not}, WebQuestions~\cite{berant2013semantic}) and two multi-hop QA datasets (HotpotQA~\cite{yang2018hotpotqa}, 2WikiMultiHopQA~\cite{ho2020constructing}). Each dataset follows the format ``\textit{question, gold answer}'', representing the user query $q$ and the gold answer $a$, respectively. The corpus $K$ used for retrieval, also referred to as the external knowledge base, is set to the Wikipedia corpus. Please note that to comprehensively evaluate the robustness of existing methods against query entry errors, we conduct both in-domain and cross-domain robustness assessments. Following~\cite{xu2024sparsecl}, we use HotpotQA as the source dataset, meaning we fine-tune the retrieval model exclusively on HotpotQA. Testing on HotpotQA constitutes in-domain evaluation while testing on other datasets represents cross-domain evaluation.

\textbf{Step 2: Query Corruption.}
We utilize the nlpaug tool~\cite{ma2019nlpaug} to inject three types of query entry errors into the six collected datasets, forming the corrupted queries:
(1) \textbf{Keyboard Proximity Errors.} When users interact with LLMs via a keyboard, mistyping may occur as a result of pressing adjacent keys. To simulate this, we replace correct letters with nearby letters on the keyboard.
(2) \textbf{Visual Similarity Errors.} When users input words through handwriting, recognition tools may misinterpret characters due to irregular handwriting or inaccurate OCR algorithms, resulting in morphological errors. To simulate these handwriting input errors, we replace correct letters with visually similar ones.
(3) \textbf{Spelling Errors.} Users may occasionally forget the correct spelling of a word and input an approximation, leading to spelling errors in the query. We simulate these errors by replacing words using a spelling error dictionary. 
Specifically, we apply a 30\% probability of selecting a word in each query, and for each selected word, a 30\% probability of corrupting a character. These probabilities reflect typical user behavior, creating a realistic test environment for RAG models.

\textbf{Step 3: Label Matching.}
Since we set relatively low probabilities for both selecting a word and corrupting a character, we assume the corruption does not affect the underlying user information need and realistically simulates typical user query behavior. Therefore, we retain the original RAG labels for the corrupted queries. In other words, for an original data sample 
$(q,a)$, we replace it with $(q',a)$ where $q'$ is the corrupted version of $q$ containing one of the three entry errors, while $a$ remains unchanged.
Additionally, to evaluate model robustness under different levels of noise, we generate two versions of the QE-RAG dataset by corrupting 20\% and 40\% of the queries, representing moderate and high-error scenarios.

\textbf{Dataset Statistics and Analysis.}
Table~\ref{tab:data_main} presents the statistical analysis of the six datasets we constructed. It can be observed that the difference in the average number of words per query between corrupted queries (with error ratios of 20\% and 40\%) and original queries is not significant. This similarity indicates that our corruption strategy effectively mirrors real-world scenarios of user query entry errors. Additionally, our corruption strategy does not alter the syntactic structure of the sentences, as shown by the minimal difference in the average query length between original and corrupted queries in Table~\ref{tab:data_main}, further ensuring the quality of our QE-RAG dataset.

\textbf{Evaluation.}
Following~\cite{jin2024flashrag}, QE-RAG support EM (Exact Match), F$_1$ (token-level F$_1$ score), and Acc (Accuracy) to evaluate the effectiveness and robustness against query entry errors of RAG methods. In this paper, we use F$_1$ for evaluation, as it better reflects the accuracy of the fine-grained information in the model's generated content. Additionally, we have developed a Python framework that facilitates the easy reproduction of experiments and the integration of new datasets and additional RAG methods.

\section{Preliminary Experiments and Methodology}

In this section, we first explore how query entry errors impact the performance of the RAG system through preliminary experiments. Then, we introduce two approaches: a contrastive learning-based robust retriever training method and a retrieval-augmented query correction method, both designed to enhance robustness against query entry errors.

\subsection{Preliminary experiments}\label{sec:pre experiments}
We conducted preliminary experiments on the HotpotQA and NQ datasets to investigate the impact of 40\% and 20\% ratio query entry errors on LLM-generated outputs when the LLMs are Llama3 and Qwen2. For this analysis, we kept the handling of correct queries unchanged and focused solely on scenarios involving corrupted queries. To evaluate the effect of various strategies for mitigating the impact of errors, we tested the following approaches: 
\begin{itemize}[leftmargin=*]
\item QE-DE (Query with Errors - Document Retrieved via Errors): The corrupted query is used to retrieve three documents (the same as below), which are then fed to the LLM for generation. This represents the baseline performance when corrupted queries are directly used without any correction. 
\item QE-DC (Query with Errors - Document Retrieved via Correct Query): The corrupted query is paired with the documents retrieved using the corresponding correct query. Both are provided to the LLM for generation. This method evaluates whether providing documents retrieved with the correct query can mitigate the negative impact of query errors. 
\item QC-DE (Corrected Query - Document Retrieved via Errors): The corrected query (corresponding to the corrupted query) is used alongside the documents retrieved using the corrupted query. This tests the effectiveness of query correction in improving LLM outputs despite inaccurate retrieval. 
\item QC-DC (Corrected Query - Document Retrieved via Correct Query): The corrected query is paired with documents retrieved using the corresponding correct query. This represents the optimal scenario, where both the query and retrieval documents are corrected, and serves as an upper bound for the performance improvements achievable by correcting queries and retrieval results.
\end{itemize}

\begin{figure}[t]
    \centering
        \subfigure[Results on HotpotQA dataset.]
    {
    \includegraphics[width=0.97 \linewidth]{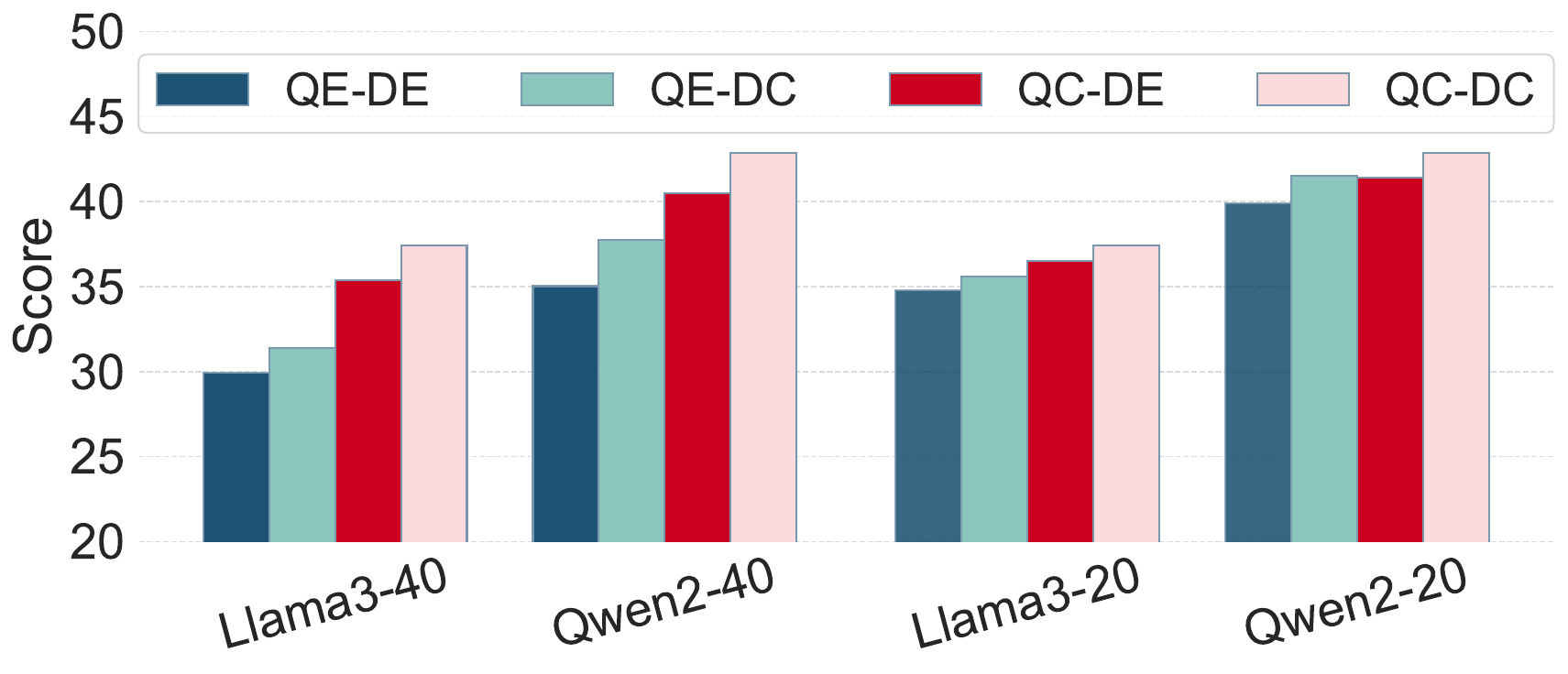}
    \label{fig:qe hotpot}
    }
    \subfigure[Results on Natural Questions dataset.]
    {
    \includegraphics[width=0.97 \linewidth]{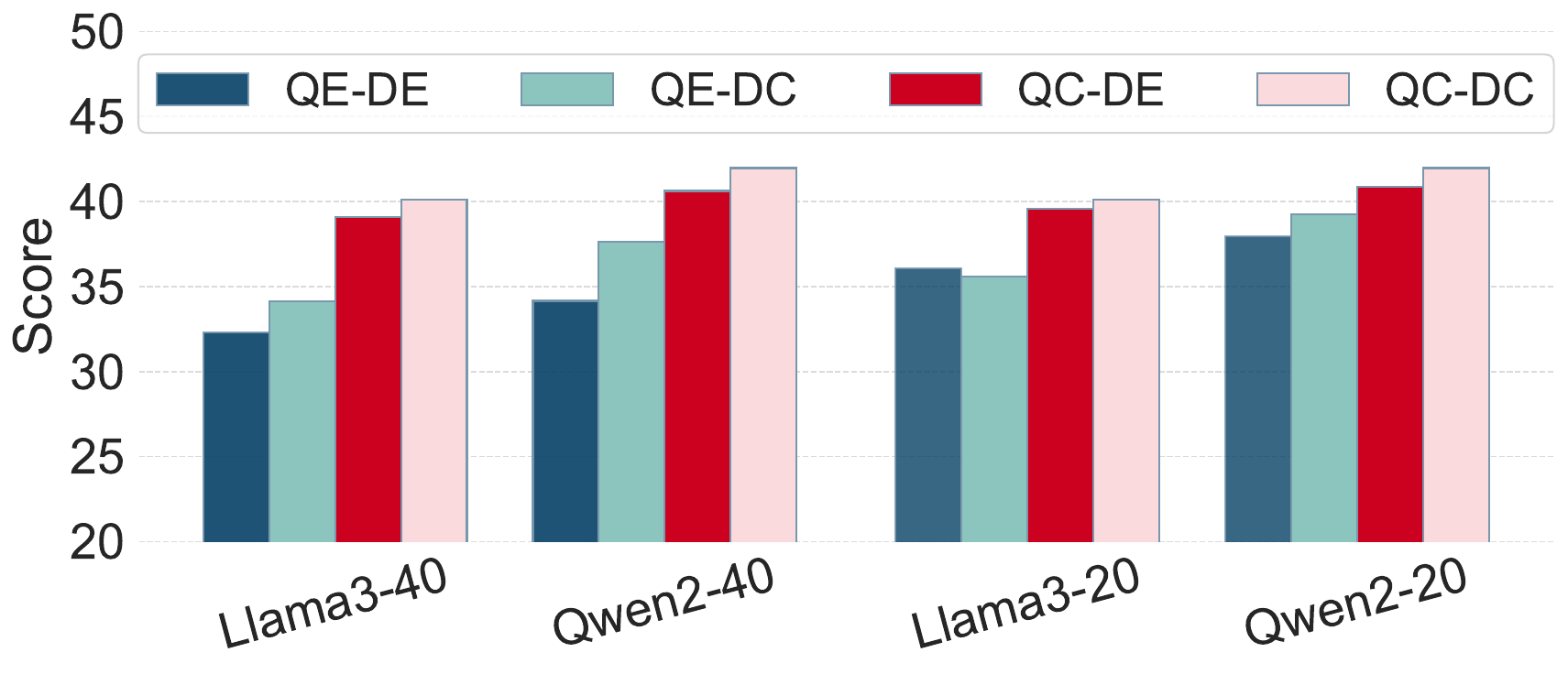}
    \label{fig:qe nq}
    }
    \caption{
    Preliminary experiments to explore the impact of query entry errors on RAG performance, where the retriever is BGE, with error ratios of 40\% and 20\%.
    }
\label{fig:intro2}
\vspace{-2mm}
\end{figure}

As shown in Figure~\ref{fig:intro2}, whether multi-hop QA (Figure~\ref{fig:intro2} (a)) or direct QA (Figure~\ref{fig:intro2} (b)), employing corrupted queries and their retrieved documents (QE-DE) gets poor model performance. In contrast, utilizing documents retrieved with correct queries (QE-DC) or using correct queries themselves (QC-DE) improved model performance. The combination of correct queries and the documents retrieved with those queries (QC-DC) achieved the best results.

Based on the above conclusions, we can infer that \textbf{retrieving correct documents for corrupted queries} and \textbf{query correction} can help address the issue of query entry errors and improve the model's robustness. Therefore, we design a contrastive learning-based robust retriever training method and a retrieval-augmented query correction method, which will be detailed in \textsection~\ref{sec:cl} and \textsection~\ref{sec:ragc}.

\subsection{\mbox{Contrastive Learning-Based Robust Retriever}}\label{sec:cl}
In order to enable the retriever to retrieve correct documents using the corrupted query, we introduce a contrastive learning-based robust retriever training method in this section.
Contrastive learning (CL) is a self-supervised learning technique designed to learn robust representations by contrasting positive and negative examples. Its advantage lies in enhancing the model's discriminative power by bringing semantically similar pairs closer in the embedding space while pushing dissimilar pairs further apart. This makes it particularly effective in scenarios where the model needs to distinguish subtle differences between inputs, such as query entry errors. Thus, we leverage CL to train the model to recognize and retrieve relevant documents even when queries are corrupted.

Specifically, we use the HotpotQA dataset, introducing a 20\% corrupted query ratio to construct contrastive pairs in the format $(q, a)$ and $(q', a)$, where $q$ and $q'$ respectively denotes the original and corrupted query, and $a$ denotes the golden LLM response. We then fine-tuned BGE~\cite{xiao2023c} models using contrastive learning on this dataset, with positive examples being the relevant documents corresponding to the original queries in HotpotQA. For negative example sampling, we included a hard negative example for each corrupted query, randomly chosen from the original HotpotQA corpus, along with randomly selected in-batch soft negative examples. The training objective is:
\begin{equation}
\label{eq:cl}
\mathcal{L}=-\mathrm{log}\frac{e^{\mathrm{sim}(\mathbf{q'_i},\mathbf{d_i^+})/\tau }}{ e^{\mathrm{sim}(\mathbf{q'_i},\mathbf{d_i^+})/\tau }+{\textstyle \sum_{j=1}^{N}}e^{\mathrm{sim}(\mathbf{q'_i},\mathbf{d_j^-})/\tau} } ,
\end{equation}
where $\mathbf{q'_i}$, $\mathbf{d_i^+}$, and $\mathbf{d_i^-}$ denote the embeddings of the 
$i$-th corrupted query, the positive example, and the negative example, respectively. The function $\mathrm{sim}(\cdot)$ represents the cosine similarity function, $N$ is the batch size, and $\tau$ is the temperature.

\subsection{Retrieval-Augmented Query Correction}\label{sec:ragc}
To better adapt to the RAG scenario, in this section, we will explore query correction using LLMs in the RAG setting.
As noted in~\cite{li2023effectiveness}, LLMs tend to overcorrect during correction tasks, modifying parts of the query that do not require changes, which may disrupt the original intent of the user. This issue arises because LLMs favor generating more common and fluent expressions, which may not align with the user's intended meaning. When user queries contain errors, the sensitivity of LLMs to prompts can exacerbate this overcorrection problem.
Our experiments also reveal that directly instructing an LLM to correct the original query often results in poor results.
In addition, for QA tasks, this behavior is problematic as LLMs may lack the necessary knowledge to provide accurate answers on their own~\cite{zhang2024trigger}. Incorporating RAG can assist LLMs in answering questions by retrieving relevant documents.
However, in the presence of query errors, providing LLMs with related documents can further complicate query correction. The LLM may prioritize answering the query based on the retrieved documents rather than focusing on the correction task. This occurs because the retrieval results may overwhelm the LLM, leading it to shift its focus from correcting the query to generating a response.
To address these challenges, we propose using retrieval-augmented fine-tuning~\cite{zhang2024raft} to efficiently fine tuning LLMs to leverage retrieved documents specifically for query correction. This approach ensures the model remains focused on correcting the query without deviating from answering it. That is:
\begin{equation}
\label{eq:raft}
\small
    \mathcal{L}_{\mathrm{FT}} = - \frac{1}{|\mathcal{D}_{I}|} \sum_{\mathcal{D}_{I}} \mathrm{log}(P_{\theta_{1}+\theta_{L}}(y_t|x,p,y_{<t})),
\end{equation}
where $\theta_{1}$ and $\theta_{L}$ are the parameters of LLM and LoRA~\cite{hu2021lora}. $y_t$ and $y_{<t}$ respectively denote the $t$-th token and tokens before $y_t$. $x$ denotes the original query with the retrieved documents. $p$ is a prompt that allows the LLM to correct the query based on the retrieved documents.  $D_I$ represents the fine-tuning dataset composed of inputs $x$, $p$ and the output, the correct query $y$.

\section{Experiments}\label{sec:experiment}
Using our QE-RAG, we aim to: (1) assess the robustness of state-of-the-art RAG methods including query reformulation methods, document refinement methods, branching and iterative methods against query entry errors; and (2) evaluate the effectiveness of our two proposed solutions: a contrastive learning-based robust retriever and a query correction approach that combines RAG with fine-tuning against these errors.
\begin{table*}[t]
    \centering
    \caption{The overall performance of the RAG task under six datasets and two different error proportions of query scenarios when the retrieval model is BGE, and the generator models are Llama3 and Qwen2. 
    The \textbf{``overall''} column represents the average result of that row, which is the average result of the method across all datasets and the two LLMs. The optimal \textbf{``overall''} results are presented in bold. 
    }
    \resizebox{1\textwidth}{!}{
    \begin{tabular}{l| cccccc| cccccc|c}
        \toprule
        \textbf{Dataset} &\textbf{HotpotQA}& \textbf{NQ} & \textbf{PopQA} &\textbf{TrivalQA}& \textbf{WebQA} & \textbf{2wiki} &\textbf{HotpotQA}& \textbf{NQ} & \textbf{PopQA} &\textbf{TrivalQA}& \textbf{WebQA} & \textbf{2wiki} &\textbf{Overall} \\ 
        \hline
        \multirow{2}{*}{\textbf{Method}}&\multicolumn{6}{c}{\textbf{Llama3}}&\multicolumn{6}{c}{\textbf{Qwen2}}&\\
        \cline{2-14}
        &\multicolumn{12}{c}{\textbf{40\% Corrupted Queries}} \\
        \hline
Standard RAG & 29.92 & 32.30 & 33.22 & 52.27 & 28.65 & 16.94 & 35.02 & 34.16 & 35.90 & 52.85 & 31.16 & 30.26 & 34.39 \\
CoT-RAG & 29.58 & 32.24 & 33.04 & 52.31 & 28.94 & 16.97 & 36.49 & 36.07 & 37.54 & 54.07 & 32.42 & 32.00 & 35.14 \\
Direct-Correct & 22.26 & 30.29 & 32.33 & 36.30 & 24.92 & 16.15 & 34.78 & 33.92 & 35.95 & 52.94 & 31.33 & 30.07 & 31.77 \\
HyDE & 7.16 & 17.82 & 2.36 & 19.58 & 12.79 & 4.35 & 25.10 & 23.33 & 29.68 & 33.21 & 22.14 & 25.07 & 18.55 \\
Iter-Retgen & 29.29 & 32.24 & 32.99 & 52.02 & 28.99 & 27.19 & 9.72 & 14.99 & 8.13 & 24.96 & 14.19 & 5.66 & 23.36 \\
REPLUG & 26.39 & 29.93 & 28.08 & 49.40 & 29.12 & 17.63 & 31.14 & 26.49 & 27.80 & 47.65 & 25.50 & 27.24 & 30.53 \\
LongLingua & 28.02 & 29.24 & 30.66 & 50.38 & 29.84 & 20.55 & 25.75 & 21.85 & 19.96 & 42.92 & 24.06 & 24.87 & 29.01 \\
SuRe & 24.50 & 32.96 & 38.42 & 47.84 & 31.35 & 14.81 & 31.48 & 27.91 & 31.02 & 51.44 & 30.48 & 28.40 & 32.55 \\
    \hdashline  
QER-RAG & 30.10 & 35.12 & 35.17 & 55.01 & 29.22 & 17.53 & 33.59 & 36.56 & 38.36 & 51.45 & 33.64 & 25.19 & 35.08 \\
RA-QCG & 31.23 & 38.44 & 35.86 & 57.87 & 30.30 & 17.80 & 38.19 & 39.04 & 38.17 & 57.00 & 33.44 & 32.98 & \textbf{37.52} \\
        \hline 
        \hline
        &\multicolumn{12}{c}{\textbf{20\% Corrupted Queries}} \\
        \hline
Standard RAG & 34.76 & 36.09 & 36.89 & 57.76 & 30.65 & 18.06 & 39.88 & 37.95 & 39.83 & 58.33 & 33.40 & 32.83 & 38.04 \\
CoT-RAG & 34.01 & 36.09 & 36.50 & 57.62 & 30.84 & 18.07 & 39.65 & 37.70 & 39.96 & 58.34 & 33.51 & 32.98 & 37.94 \\
Direct-Correct & 23.59 & 31.57 & 31.31 & 34.60 & 24.85 & 15.84 & 25.12 & 23.64 & 30.84 & 32.95 & 22.84 & 25.51 & 26.89 \\
HyDE & 7.28 & 18.76 & 2.20 & 20.32 & 11.47 & 4.27 & 9.85 & 15.79 & 7.66 & 26.61 & 15.87 & 5.80 & 12.16 \\
Iter-Retgen & 34.35 & 35.80 & 36.79 & 57.34 & 30.98 & 17.16 & 35.65 & 28.40 & 31.06 & 53.16 & 26.96 & 29.60 & 34.77 \\
REPLUG & 30.24 & 33.55 & 31.55 & 54.27 & 31.26 & 20.27 & 28.89 & 25.43 & 22.25 & 47.05 & 26.46 & 26.93 & 31.51 \\
LongLingua & 33.30 & 33.43 & 32.96 & 56.94 & 31.42 & 23.21 & 34.74 & 31.58 & 34.26 & 55.59 & 33.14 & 31.23 & 35.98 \\
 SuRe & 27.91 & 36.97 & 42.17 & 53.17 & 34.81 & 16.19 & 38.23 & 40.31 & 43.78 & 56.29 & 35.54 & 27.78 & 37.76 \\
    \hdashline  
QER-RAG & 33.31 & 38.24 & 38.71 & 58.85 & 31.83 & 18.95 & 39.84 & 39.21 & 41.43 & 58.66 & 34.83 & 35.24 & 39.09 \\
RA-QCG & 35.08 & 39.64 & 39.02 & 60.55 & 32.26 & 19.62 & 41.65 & 40.71 & 41.84 & 59.77 & 35.74 & 36.03 & \textbf{40.16} \\

        \bottomrule
    \end{tabular}
    }

    \label{tab:bge main}
\vspace{-3mm}
\end{table*}

\subsection{Experimental Settings}

\subsubsection{Datasets and Metrics}
In the main experiment, we selected our modified RAG dataset to conduct experiments on RAG tasks. Specifically, we chose four QA datasets: TriviaQA~\cite{joshi2017triviaqa}, Natural Questions (NQ)~\cite{kwiatkowski2019natural}, PopQA~\cite{mallen2022not}, WebQuestions (WebQA)~\cite{berant2013semantic}, and two Multi-Hop QA datasets: HotpotQA~\cite{yang2018hotpotqa} and 2WikiMultihopQA(2wiki)~\cite{ho2020constructing} for our experiments. 
Following~\cite{jin2024flashrag}, we used the Wikipedia data from December 2018 as the retrieval corpus.
For the evaluation metrics, following~\cite{jin2024flashrag}, we selected the widely used token-level F$_1$ score as our evaluation metric. We also support the use of other evaluation metrics.

\subsubsection{Retrieval and Generation Models}
In our main experiment, we selected the sentence embedding models with SOTA performance on the MTEB leaderboard~\cite{muennighoff2022mteb}, namely bge-base-en-v1.5~\cite{xiao2023c} as the retrieval models. Other retrieval models can also be adapted to our benchmark. 
As \textsection~\ref{sec:cl} described, We trained them on the original HotpotQA dataset as well as the HotpotQA dataset we constructed with 20\% corrupted queries, obtaining retrievers R$_1$ and R$_2$ respectively. For the baseline, we used R$_1$ as the retriever. For our method, we used R$_2$ as the retriever.
For the generation models, we chose the latest Llama3-8B-Instruct~\cite{llama3modelcard} and Qwen2-7B-Instruct~\cite{yang2024qwen2} as the main experimental generation models. They have been proven to have strong performance in RAG tasks. 

\begin{table}[t]
    \centering
        \caption{The compatibility with existing RAG methods when the error rate is 20\% and the LLM is Llama3.
    }
    \resizebox{1\linewidth}{!}{
    \begin{tabular}{l| cccccc}
        \toprule
        \textbf{Method} &\textbf{HotpotQA}& \textbf{NQ} & \textbf{PopQA} &\textbf{TrivalQA}& \textbf{WebQA} & \textbf{2wiki} \\ 
        \hline
        Iter-Retgen & 34.35 & 35.80 & 36.79 & 57.34 & 30.98 & 17.16  \\
+RA-QCG & \textbf{35.45} & \textbf{38.94} & \textbf{38.94} & \textbf{60.84} & \textbf{32.71} & \textbf{19.27}   \\
\hline
REPLUG & 30.24 & 33.55 & 31.55 & 54.27 & 31.26 & 20.27  \\
+RA-QCG & \textbf{31.19} & \textbf{36.25} & \textbf{34.53} & \textbf{58.78} & \textbf{32.86} & \textbf{22.01}   \\
\hline
LongLingua & 33.30 & 33.43 & 32.96 & 56.94 & 31.42 & 23.21   \\
+RA-QCG & 32.76 & \textbf{35.56} & \textbf{34.22} & \textbf{58.15} & \textbf{32.92} & \textbf{23.50}   \\
\hline
 SuRe & 27.91 & 36.97 & 42.17 & 53.17 & 34.81 & 16.19   \\
+RA-QCG & \textbf{29.41} & \textbf{39.12} & \textbf{44.28} & \textbf{55.21} & \textbf{35.91} & \textbf{18.33}  \\
        \bottomrule
    \end{tabular}
    }

    \label{tab:bge rag methods}
\end{table}

\begin{table*}[t]
    \centering
    \caption{The overall performance of the RAG task under six datasets and 0\% error proportions of query scenarios when the retrieval model is BGE, and the generator models are Llama3 and Qwen2. 
    The \textbf{``overall''} column represents the average result of that row, which is the average result of the method across all datasets and the two LLMs. The optimal \textbf{``overall''} results are presented in bold. 
    }
    \resizebox{1\textwidth}{!}{
    \begin{tabular}{l| cccccc| cccccc|c}
        \toprule
        \textbf{Dataset} &\textbf{HotpotQA}& \textbf{NQ} & \textbf{PopQA} &\textbf{TrivalQA}& \textbf{WebQA} & \textbf{2wiki} &\textbf{HotpotQA}& \textbf{NQ} & \textbf{PopQA} &\textbf{TrivalQA}& \textbf{WebQA} & \textbf{2wiki} &\textbf{Overall} \\ 
        \hline
        \multirow{2}{*}{\textbf{Method}}&\multicolumn{6}{c}{\textbf{Llama3}}&\multicolumn{6}{c}{\textbf{Qwen2}}&\\
        \cline{2-14}
        &\multicolumn{12}{c}{\textbf{0\% Corrupted Queries}} \\
        \hline
Standard RAG & 37.40 & 40.10 & 40.83 & 63.32 & 33.56 & 20.72 & 42.87 & 41.97 & 43.66 & 64.44 & 36.74 & 36.49 & 41.84 \\
CoT-RAG & 36.84 & 40.03 & 40.44 & 63.07 & 33.95 & 20.68 & 42.52 & 41.94 & 43.84 & 64.46 & 36.98 & 36.31 & 41.76 \\
Direct-Correct & 23.14 & 33.22 & 37.53 & 33.99 & 27.22 & 16.80 & 26.02 & 23.39 & 32.64 & 31.74 & 22.64 & 26.81 & 27.93 \\
HyDE & 8.06 & 19.92 & 2.27 & 22.08 & 12.12 & 4.93 & 9.99 & 16.51 & 7.43 & 28.11 & 17.12 & 5.48 & 12.83 \\
 Iter-Retgen & 36.81 & 39.82 & 40.49 & 63.08 & 33.78 & 19.37 & 38.84 & 31.76 & 34.42 & 58.69 & 26.03 & 32.84 & 37.99 \\
REPLUG & 33.83 & 37.53 & 34.39 & 59.99 & 34.98 & 21.83 & 32.42 & 28.60 & 24.45 & 52.71 & 28.78 & 29.52 & 34.92 \\
LongLingua & 35.47 & 37.31 & 36.14 & 61.88 & 34.77 & 25.92 & 38.18 & 35.69 & 37.90 & 61.15 & 35.33 & 34.51 & 39.52 \\
SuRe & 30.20 & 41.54 & 46.12 & 59.12 & 39.17 & 19.10 & 42.09 & 44.43 & 48.70 & 62.74 & 39.92 & 31.60 & 42.06 \\
\hdashline
QER-RAG & 36.32 & 41.55 & 41.59 & 63.67 & 34.45 & 20.69 & 42.92 & 42.73 & 44.57 & 63.70 & 37.70 & 38.09 & \textbf{42.33} \\
RA-QCG & 36.22 & 41.50 & 41.59 & 63.70 & 34.51 & 20.68 & 42.90 & 42.73 & 44.57 & 63.71 & 37.77 & 38.09 & \textbf{42.33} \\
        \bottomrule
    \end{tabular}
    }

    \label{tab:bge ration0}
\end{table*}

\subsubsection{RAG Methods}
We test the following RAG methods.
We begin by evaluating the \textbf{Standard RAG} method, where the LLM generates responses directly based on the retrieved documents. We extend this baseline by introducing \textbf{CoT-RAG}, which prompts the LLM to consider whether the original query contains errors while generating a response.
For query reformulation baselines, we focus on cost-effective, training-free approaches for evaluation: \textbf{Direct-Correct}: The LLM corrects the input query directly, and the corrected query is used for retrieval. \textbf{HyDE}~\cite{gao2023precise}: The LLM generates a pseudo-document answering the query, which is then used as the new query for retrieval. \textbf{Iter-Retgen}~\cite{shao2023enhancing}: This method iteratively refines retrieval by leveraging the LLM’s responses combined with the original query as new retrieval queries.
To evaluate methods that refine retrieved documents, we consider \textbf{LongLingua}~\cite{jiang2023longllmlingua}, which uses the LLM to modify the retrieved documents based on the query perplexity, making them more interpretable and better aligned with the LLM’s contextual understanding.
For branching methods, we evaluate: \textbf{REPLUG}~\cite{shi2024replug}: Enhances response generation by integrating document relevance into the output. \textbf{SuRe}~\cite{kimsure}: Summarizes multiple candidate answers to determine the most appropriate response.
All the above methods use R1 as the retriever.
For our proposed methods: \textbf{QER-RAG}: To enhance the robustness of retrieval, we replace the retriever R1 with our trained retriever R2 while keeping other components of the standard RAG method unchanged. \textbf{RA-QCG}: This method integrates our query correction approach into standard RAG. The original query is corrected using retrieved documents, and the corrected query is then used for RAG.

\subsubsection{Implementation Details}
We employed the HuggingFace Transformers~\cite{wolf2020transformers} in PyTorch for the experiments. 
We set the generation parameter do\_sample to false to improve the reproducibility of the results.
Except for the experiment in \textsection~\ref{sec:nums} on the impact of the number of retrievals on robustness, in all RAG tasks, three documents are retrieved for each query given the computational costs.
We set the maximum input length to 4096 for the generation models. Following~\cite{jin2024flashrag}, we test 1000 queries for each RAG dataset.
For the training of contrastive learning models in \textsection~\ref{sec:cl}, we set the learning rate to 2e-5, batch size to 64, and epoch to 1.
We use LoRA~\cite{hu2021lora}  for efficient fine-tuning of LLMs, using the Adam optimizer~\cite{kingma2014adam}, setting the initial learning rate to 5e-5, batch size to 16, and employing a cosine learning rate schedule. We train for 3 epochs with 1,000 pieces of data from the training dataset of HotpotQA with a 20\% error rate.
For Iter-Retgen, we iterate one round. For LongLingua, we use LLM itself as the compressor, with the compression rate set to 0.5 and the rest consistent with the original paper. For REPLUG, we keep its original settings. For SuRe, we use the prompt provided in the original paper to summarize and select candidate answers.
All experiments are conducted on Nvidia A6000 GPUs. More details can be found at the link provided in the Evaluation part of \textsection~\ref{sec:main sec}.

\vspace{-2mm}
\subsection{Main Results}\label{sec:main res}
Table~\ref{tab:bge main} shows the main experimental results of different methods in six QE-RAG datasets with two different corrupted query proportions (20\%, 40\%) when the retrieval model is BGE.
From the table, we can draw the following observations:

\textbf{The Poor Robustness of Existing SOTA RAG Methods.}
It can be observed that when the dataset contains corrupted queries (with error ratios of 20\% or 40\%), the performance of existing SOTA RAG methods in performing is suboptimal. As the proportion of corrupted queries increases, the model's performance deteriorates progressively, indicating its lack of robustness when handling query entry errors. This phenomenon underscores the critical importance of handling query entry errors for the success of RAG tasks. 
Despite many SOTA methods optimizing RAG components and employing various strategies such as query reformulation, compressing retrieval documents to improve LLM comprehension, handling different cases through branching, or using iterative generation to enhance retrieval, they still struggle to effectively handle query entry errors. The reason is that if the original query is corrupted, it may confuse the model, preventing it from correctly understanding the task and leading to incorrect answers. For example, in the case of HyDE, if the LLM is asked to generate a response document based on the corrupted query, it may result in even more severe errors because the LLM may not understand the corrupted query in the first place. Therefore, addressing the query entry errors in RAG scenarios is crucial to ensure that the model provides more accurate and reliable answers, thereby enhancing the user experience.

\textbf{The Effectiveness of QER-RAG.}
Our proposed QER-RAG method builds upon the standard RAG with improvements. Specifically, QER-RAG differs from standard RAG in that it uses a retriever trained on a dataset containing corrupted queries. Experimental results show that QER-RAG achieves significant improvements at both error ratios (20\% and 40\%). This result demonstrates the effectiveness of the contrastive learning approach we introduced in training the retriever with a dataset containing corrupted queries. By incorporating a certain proportion (specifically, 20\%) of corrupted queries into the retriever's training data, we can significantly improve the retriever's robustness, allowing it to still retrieve relevant documents in the face of corrupted inputs and helping the LLM generate more accurate responses. In addition, the slight decrease in certain in-domain metrics may be due to the imbalance in the data samples.

\textbf{The Effectiveness of RA-QCG.}
Building on QER-RAG, we further propose the RA-QCG method, which introduces a query correction mechanism based on RAG. Experimental results show that RA-QCG achieves optimal overall performance at both error ratios (20\% and 40\%), and in the case of a 40\% error ratio, RA-QCG's performance even approaches the best baseline performance observed at the 20\% error ratio. This result fully validates the effectiveness of our RAG-assisted query correction approach. Compared to traditional query reformulation, iterative retrieval, document compression, and other methods, our approach significantly improves RAG performance without increasing the number of LLM calls. This shows that RA-QCG achieves superior RAG performance through query correction without adding additional computational overhead.

\textbf{Dataset-Specific Findings.}
The experimental results indicate that in scenarios involving corrupted queries, SOTA RAG methods do not always outperform the standard RAG method, especially on certain specific datasets. This finding is consistent with conclusions from~\cite{jin2024flashrag} where, in the absence of corrupted queries, SOTA RAG methods do not always perform optimally. The reasons for this can be attributed to two main issues: first, the failure to retrieve relevant documents can prevent SOTA RAG methods from fully leveraging their strengths. This is often because the retrieval corpus (Wikipedia data from December 2018) may not cover the answers to the questions, or the retrieval model may not be powerful enough. Second, retrieving irrelevant documents introduces noise into the LLM generation process, and since LLMs are highly sensitive to prompts, this noise can negatively impact the quality of the generated results.

\vspace{-2mm}
\subsection{Compatibility with SOTA RAG}
From the main experiments in \textsection~\ref{sec:main res}, we observe that state-of-the-art RAG methods offer notable improvements over standard RAG methods. This inspired us to explore whether our proposed approach is compatible with these methods, potentially further enhancing RAG system performance and robustness.
In this section, we investigate the effectiveness of combining our method with four advanced RAG methods—IterGen, LongLingua, RePlug, and Sure—under the setting where the LLM is LLama3 and the query error rate is 20\%. These methods represent a diverse range of strategies.

The results are shown in Table~\ref{tab:bge rag methods}.
The performance gains from our method are observed across all tested RAG methods, demonstrating its generalizability and flexibility in complementing diverse retrieval and reasoning strategies.
By incorporating our query correction mechanism and robust retrieval approach, these methods show enhanced robustness when handling queries with entry errors.
The results underscore that our method is not only effective as a standalone solution but also as an enhancement to existing SOTA RAG approaches. Its compatibility with SOTA methods allows it to serve as a modular addition to RAG pipelines, making it a valuable tool for building robust and high-performing retrieval-augmented generation systems.

\subsection{Robustness on Correct Queries}\label{sec:ration-0}
In this section, we investigate the robustness of our proposed method when the query error rate is 0\%. Specifically, we aim to assess whether focusing on handling corrupted queries negatively impacts performance on correct queries. For this evaluation, we use the same models and RAG methods as in the main experiments, but the dataset consists entirely of correct queries.

The results, presented in Table~\ref{tab:bge ration0}, demonstrate that our method achieves the best overall performance when all queries are correct. This highlights the robustness of our approach, which does not compromise its ability to handle correct queries despite its emphasis on addressing corrupted queries.
Additionally, comparing Table~\ref{tab:bge main} with Table~\ref{tab:bge ration0} reveals that the performance of all methods improves when the queries are error-free. This observation further validates the findings from our preliminary experiments in \textsection~\ref{sec:pre experiments}: correcting query entry errors such as keyboard proximity errors, visual similarity errors, and spelling mistakes can enhance the overall performance of RAG systems. By improving the accuracy and relevance of retrieved documents, such corrections contribute to a better user experience.
Overall, these results confirm that our method effectively balances robustness across both corrupted and correct queries, ensuring high performance in real-world scenarios where query quality varies.

\subsection{Robustness Comparison of Correct and Corrupted Query}\label{sec:part}

\begin{figure}[t]
    \centering
\includegraphics[width=\linewidth]{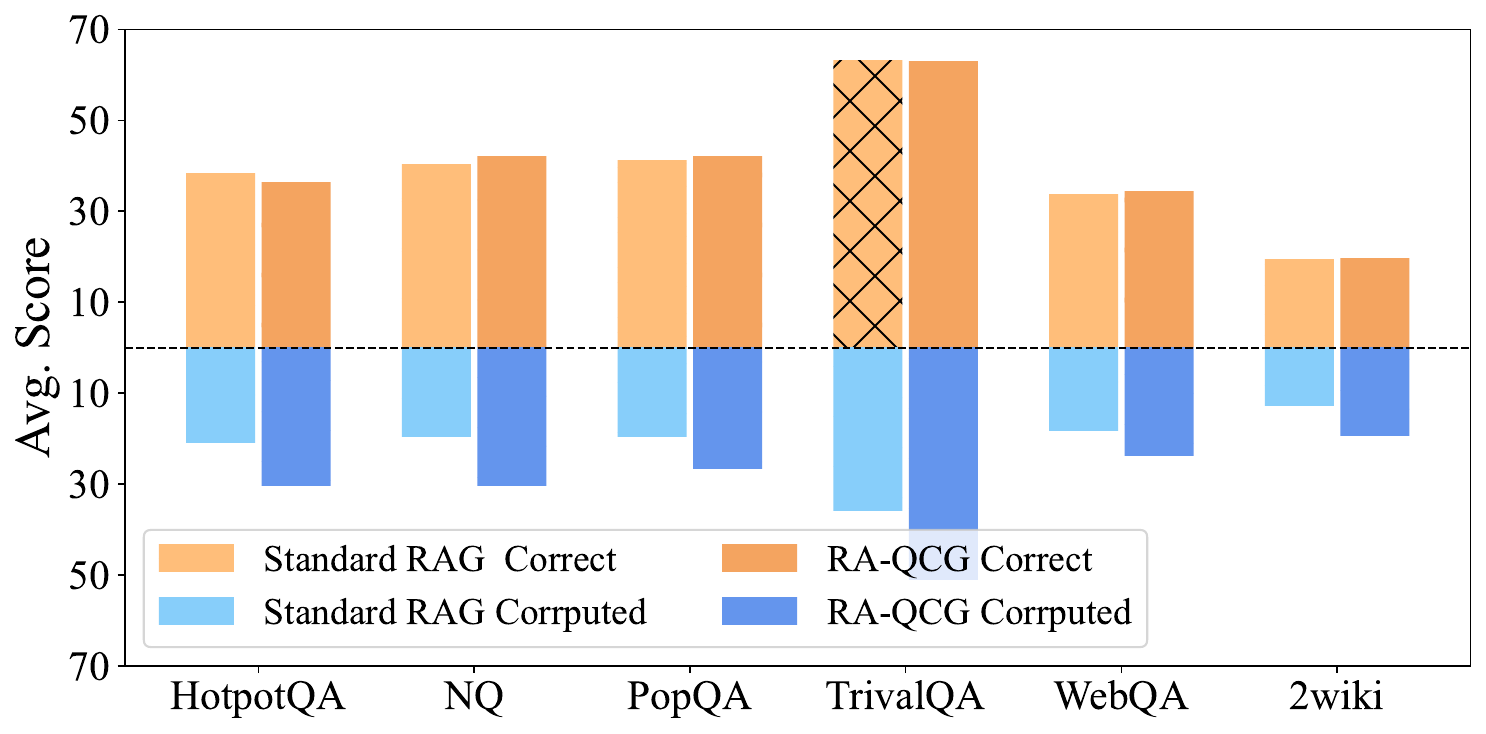  }
    \caption{
    The robustness comparison of correct and corrupted queries to the average F$_1$ score when the retrieval model is Standard RAG and RA-QCG, the generative model is Llama3 and the error rate is 20\%. 
    Above and below the X-axis represent the average token level F$_1$ value of the correct and corrupted query, respectively.
    }
    \label{fig:part}
    \vspace{-3mm}
\end{figure}

Table~\ref{tab:bge main} in the main experiment shows the overall RAG performance for all queries (correct and corrupted queries), but we are unaware of how the RAG model performs on correct versus corrupted queries individually. RA-QGC improves upon standard RAG. Therefore, in this section, we explore the average F$_1$ scores of RA-QGC and standard RAG across six datasets with a 20\% corrupted query ratio when the LLM is Llama3. We have a total of 1000 queries, of which 200 are corrupted and 800 are correct. 

The results are shown in Figure~\ref{fig:part}.
It can be seen that the average performance on correct queries is similar across all six datasets, while for corrupted queries, RA-QGC demonstrates a significant advantage, with its average score outperforming standard RAG across all datasets. In addition to \textsection~\ref{sec:ration-0}, this experiment further illustrates that RA-QGC can effectively improve the robustness of the RAG method in both in-domain and cross-domain datasets when faced with query entry errors, thus enhancing the overall performance of the RAG method.

\begin{figure}[t]
    \centering
\includegraphics[width=\linewidth]{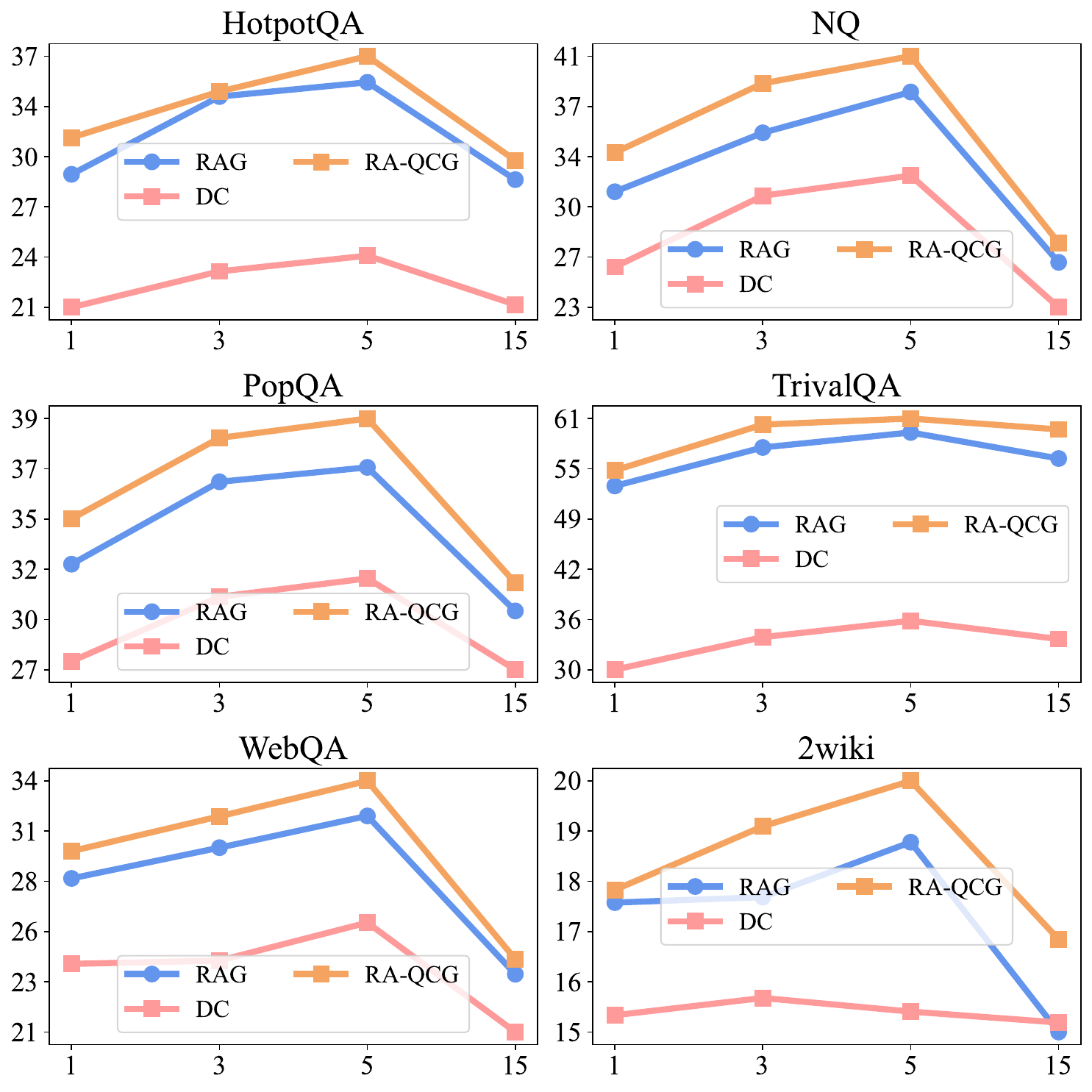}
    \caption{
    The results of Standard RAG (RAG), Direct-Correct (DC) and  RA-QCG retrieving varying numbers of documents on six datasets when the LLM is Llama3 and the error rate is 20\%.
    The x-axis represents the number of retrieved documents, specifically 1, 3, 5, and 15, while the y-axis indicates the token-level F$_1$ score.
    }
    \label{fig:nums}
    \vspace{-3mm}
\end{figure}

\subsection{Robustness on the Number of Documents Retrieved}\label{sec:nums}
In RAG tasks without corrupted queries, retrieving different numbers of documents can have varying effects on RAG performance~\cite{jin2024flashrag}. When fewer documents are retrieved, the generation model may struggle to find the correct answer to the query. On the other hand, retrieving too many documents may overwhelm the generation model with excessive noise, making it difficult to focus on the key information. Therefore, in this section, we explore the impact of retrieving different numbers of documents on the robustness of RAG methods.
We test standard RAG, Direct-Correct, and RA-QGC with Llama3 as the LLM, using retrievals of 1, 3, 5, and 15 documents to supplement the LLM’s knowledge. The results are shown in Figure~\ref{fig:nums}.

The following conclusions can be drawn:
(a) Regardless of the number of documents retrieved, RA-QGC consistently achieves improvements. This indicates that RA-QGC is more robust and is not limited by the number of retrieved documents, meaning it works effectively across various resource configurations (retrieving different numbers of documents).
(b) The performance of RAG increases and then decreases as the number of retrieved documents changes, similar to the pattern observed in correct query scenarios~\cite{jin2024flashrag}. This suggests that selecting an appropriate number of documents for retrieval is crucial, balancing resources and RAG performance while accounting for the potential noise introduced by more documents.
(c) It can be seen that using LLM-based direct correction significantly worsens RAG performance, and increasing the number of retrieved documents does little to alleviate the over-correction issue in LLMs. This highlights the necessity of query correction based on RAG, which leads to more accurate corrections and, as a result, improved RAG performance.

\subsection{Qualitative Analysis on Robustness}
\begin{figure}[t]
    \centering
\includegraphics[width=\linewidth]{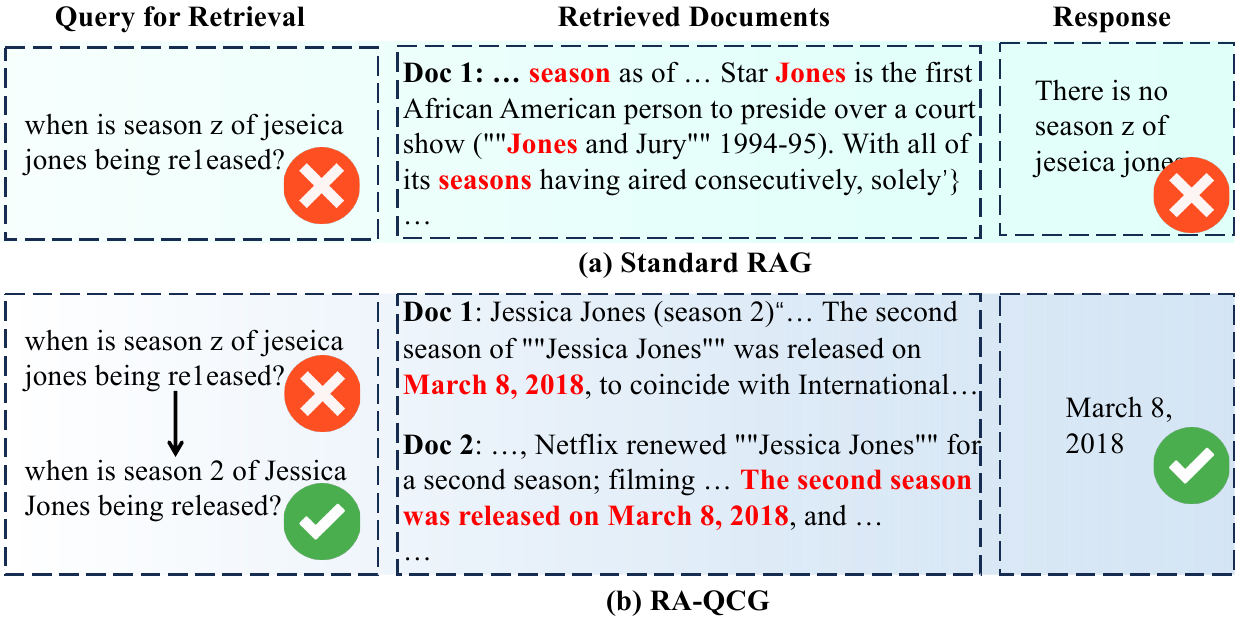}
    \caption{
    The case study of Standard RAG and  RA-QCG.
    }
    \label{fig:case study}
    \vspace{-3mm}
\end{figure}

To investigate how our proposed method enhances model robustness, we conduct a qualitative analysis. Given that our method builds on the standard RAG, we compare the performance of RA-QCG with the standard RAG using a randomly selected example from the NQ dataset, with Llama3 as the LLM. This analysis examines three key components: the query used for retrieval, the documents retrieved, and the final responses generated by the LLM. 

The results are illustrated in Figure~\ref{fig:case study}.
\textbf{Query for Retrieval.}
In standard RAG, the corrupted query provided by the user is directly used for retrieval. In contrast, RA-QCG identifies and corrects the errors in the query before the retrieval stage, effectively mitigating the impact of input inaccuracies. This step ensures that the subsequent retrieval process operates on a more accurate representation of the user's intent.
\textbf{Retrieved Documents.}
Due to the use of the corrupted query, the standard RAG retrieves documents that are misaligned with the user's intended question. As a result, the retrieved documents lack the necessary information to answer the query correctly. Conversely, RA-QCG, by utilizing the corrected query, retrieves documents that are well-aligned with the user's intent, containing the relevant information needed to address the query effectively.
\textbf{Response.}
The shortcomings of the standard RAG are evident in the response generation stage. The misaligned documents retrieved by it lead to an incoherent or incorrect response that fails to answer the user’s question. On the other hand, RA-QCG benefits from the corrected query and the retrieval of relevant documents, enabling the LLM to generate a response that is accurate and contextually appropriate.
This analysis highlights how RA-QCG successfully corrects the query, retrieves documents that provide the necessary context and produces accurate answers. RA-QCG improves the robustness and reliability of the RAG system.

\section{Conclusion}
In this paper, we present the first comprehensive investigation into the robustness of retrieval-augmented generation against query entry errors. We build the QE-RAG by simulating three types of query errors: "keyboard proximity, visual similarity, and spelling" based on six RAG datasets with varying error ratios. We find that corrupted queries lead to a performance drop in the RAG methods, but this can be alleviated through query correction and retrieval model adjustments.
Based on QE-RAG, we test standard RAG, existing SOTA RAG methods (including query reformulation, document compression, branching, and iterative methods), as well as our proposed robust retrieval method, which is trained using contrastive learning on corrupted queries and retrieval-augmented query correction method. The results show that existing RAG methods exhibit poor robustness to query entry errors, while our two proposed methods effectively enhance the robustness of the RAG methods.


\bibliographystyle{ACM-Reference-Format}
\bibliography{sample-base}


\begin{thebibliography}{40}


\ifx \showCODEN    \undefined \def \showCODEN     #1{\unskip}     \fi
\ifx \showDOI      \undefined \def \showDOI       #1{#1}\fi
\ifx \showISBNx    \undefined \def \showISBNx     #1{\unskip}     \fi
\ifx \showISBNxiii \undefined \def \showISBNxiii  #1{\unskip}     \fi
\ifx \showISSN     \undefined \def \showISSN      #1{\unskip}     \fi
\ifx \showLCCN     \undefined \def \showLCCN      #1{\unskip}     \fi
\ifx \shownote     \undefined \def \shownote      #1{#1}          \fi
\ifx \showarticletitle \undefined \def \showarticletitle #1{#1}   \fi
\ifx \showURL      \undefined \def \showURL       {\relax}        \fi
\providecommand\bibfield[2]{#2}
\providecommand\bibinfo[2]{#2}
\providecommand\natexlab[1]{#1}
\providecommand\showeprint[2][]{arXiv:#2}

\bibitem[Ai et~al\mbox{.}(2023)]%
        {ai2023information}
\bibfield{author}{\bibinfo{person}{Qingyao Ai}, \bibinfo{person}{Ting Bai}, \bibinfo{person}{Zhao Cao}, \bibinfo{person}{Yi Chang}, \bibinfo{person}{Jiawei Chen}, \bibinfo{person}{Zhumin Chen}, \bibinfo{person}{Zhiyong Cheng}, \bibinfo{person}{Shoubin Dong}, \bibinfo{person}{Zhicheng Dou}, \bibinfo{person}{Fuli Feng}, {et~al\mbox{.}}} \bibinfo{year}{2023}\natexlab{}.
\newblock \showarticletitle{Information retrieval meets large language models: a strategic report from chinese ir community}.
\newblock \bibinfo{journal}{\emph{AI Open}}  \bibinfo{volume}{4} (\bibinfo{year}{2023}), \bibinfo{pages}{80--90}.
\newblock


\bibitem[AI@Meta(2024)]%
        {llama3modelcard}
\bibfield{author}{\bibinfo{person}{AI@Meta}.} \bibinfo{year}{2024}\natexlab{}.
\newblock \showarticletitle{Llama 3 Model Card}.
\newblock  (\bibinfo{year}{2024}).
\newblock
\urldef\tempurl%
\url{https://github.com/meta-llama/llama3/blob/main/MODEL_CARD.md}
\showURL{%
\tempurl}


\bibitem[Berant et~al\mbox{.}(2013)]%
        {berant2013semantic}
\bibfield{author}{\bibinfo{person}{Jonathan Berant}, \bibinfo{person}{Andrew Chou}, \bibinfo{person}{Roy Frostig}, {and} \bibinfo{person}{Percy Liang}.} \bibinfo{year}{2013}\natexlab{}.
\newblock \showarticletitle{Semantic parsing on freebase from question-answer pairs}. In \bibinfo{booktitle}{\emph{Proceedings of the 2013 conference on empirical methods in natural language processing}}. \bibinfo{pages}{1533--1544}.
\newblock


\bibitem[Borgeaud et~al\mbox{.}(2022)]%
        {borgeaud2022improving}
\bibfield{author}{\bibinfo{person}{Sebastian Borgeaud}, \bibinfo{person}{Arthur Mensch}, \bibinfo{person}{Jordan Hoffmann}, \bibinfo{person}{Trevor Cai}, \bibinfo{person}{Eliza Rutherford}, \bibinfo{person}{Katie Millican}, \bibinfo{person}{George~Bm Van Den~Driessche}, \bibinfo{person}{Jean-Baptiste Lespiau}, \bibinfo{person}{Bogdan Damoc}, \bibinfo{person}{Aidan Clark}, {et~al\mbox{.}}} \bibinfo{year}{2022}\natexlab{}.
\newblock \showarticletitle{Improving language models by retrieving from trillions of tokens}. In \bibinfo{booktitle}{\emph{International conference on machine learning}}. PMLR, \bibinfo{pages}{2206--2240}.
\newblock


\bibitem[Chen et~al\mbox{.}(2024)]%
        {chen2024benchmarking}
\bibfield{author}{\bibinfo{person}{Jiawei Chen}, \bibinfo{person}{Hongyu Lin}, \bibinfo{person}{Xianpei Han}, {and} \bibinfo{person}{Le Sun}.} \bibinfo{year}{2024}\natexlab{}.
\newblock \showarticletitle{Benchmarking large language models in retrieval-augmented generation}. In \bibinfo{booktitle}{\emph{Proceedings of the AAAI Conference on Artificial Intelligence}}, Vol.~\bibinfo{volume}{38}. \bibinfo{pages}{17754--17762}.
\newblock


\bibitem[Dai et~al\mbox{.}(2024)]%
        {dai2024neural}
\bibfield{author}{\bibinfo{person}{Sunhao Dai}, \bibinfo{person}{Yuqi Zhou}, \bibinfo{person}{Liang Pang}, \bibinfo{person}{Weihao Liu}, \bibinfo{person}{Xiaolin Hu}, \bibinfo{person}{Yong Liu}, \bibinfo{person}{Xiao Zhang}, \bibinfo{person}{Gang Wang}, {and} \bibinfo{person}{Jun Xu}.} \bibinfo{year}{2024}\natexlab{}.
\newblock \showarticletitle{Neural retrievers are biased towards llm-generated content}. In \bibinfo{booktitle}{\emph{Proceedings of the 30th ACM SIGKDD Conference on Knowledge Discovery and Data Mining}}. \bibinfo{pages}{526--537}.
\newblock


\bibitem[Es et~al\mbox{.}(2024)]%
        {es2024ragas}
\bibfield{author}{\bibinfo{person}{Shahul Es}, \bibinfo{person}{Jithin James}, \bibinfo{person}{Luis~Espinosa Anke}, {and} \bibinfo{person}{Steven Schockaert}.} \bibinfo{year}{2024}\natexlab{}.
\newblock \showarticletitle{RAGAs: Automated Evaluation of Retrieval Augmented Generation}. In \bibinfo{booktitle}{\emph{Proceedings of the 18th Conference of the European Chapter of the Association for Computational Linguistics: System Demonstrations}}. \bibinfo{pages}{150--158}.
\newblock


\bibitem[Fang et~al\mbox{.}(2023)]%
        {fang2023chatgpt}
\bibfield{author}{\bibinfo{person}{Tao Fang}, \bibinfo{person}{Shu Yang}, \bibinfo{person}{Kaixin Lan}, \bibinfo{person}{Derek~F Wong}, \bibinfo{person}{Jinpeng Hu}, \bibinfo{person}{Lidia~S Chao}, {and} \bibinfo{person}{Yue Zhang}.} \bibinfo{year}{2023}\natexlab{}.
\newblock \showarticletitle{Is chatgpt a highly fluent grammatical error correction system? a comprehensive evaluation}.
\newblock \bibinfo{journal}{\emph{arXiv preprint arXiv:2304.01746}} (\bibinfo{year}{2023}).
\newblock


\bibitem[Gao et~al\mbox{.}(2023a)]%
        {gao2023precise}
\bibfield{author}{\bibinfo{person}{Luyu Gao}, \bibinfo{person}{Xueguang Ma}, \bibinfo{person}{Jimmy Lin}, {and} \bibinfo{person}{Jamie Callan}.} \bibinfo{year}{2023}\natexlab{a}.
\newblock \showarticletitle{Precise Zero-Shot Dense Retrieval without Relevance Labels}. In \bibinfo{booktitle}{\emph{Proceedings of the 61st Annual Meeting of the Association for Computational Linguistics (Volume 1: Long Papers)}}. \bibinfo{pages}{1762--1777}.
\newblock


\bibitem[Gao et~al\mbox{.}(2023b)]%
        {gao2023retrieval}
\bibfield{author}{\bibinfo{person}{Yunfan Gao}, \bibinfo{person}{Yun Xiong}, \bibinfo{person}{Xinyu Gao}, \bibinfo{person}{Kangxiang Jia}, \bibinfo{person}{Jinliu Pan}, \bibinfo{person}{Yuxi Bi}, \bibinfo{person}{Yi Dai}, \bibinfo{person}{Jiawei Sun}, {and} \bibinfo{person}{Haofen Wang}.} \bibinfo{year}{2023}\natexlab{b}.
\newblock \showarticletitle{Retrieval-augmented generation for large language models: A survey}.
\newblock \bibinfo{journal}{\emph{arXiv preprint arXiv:2312.10997}} (\bibinfo{year}{2023}).
\newblock


\bibitem[Ho et~al\mbox{.}(2020)]%
        {ho2020constructing}
\bibfield{author}{\bibinfo{person}{Xanh Ho}, \bibinfo{person}{Anh-Khoa~Duong Nguyen}, \bibinfo{person}{Saku Sugawara}, {and} \bibinfo{person}{Akiko Aizawa}.} \bibinfo{year}{2020}\natexlab{}.
\newblock \showarticletitle{Constructing A Multi-hop QA Dataset for Comprehensive Evaluation of Reasoning Steps}. In \bibinfo{booktitle}{\emph{Proceedings of the 28th International Conference on Computational Linguistics}}. \bibinfo{pages}{6609--6625}.
\newblock


\bibitem[Hu et~al\mbox{.}(2021)]%
        {hu2021lora}
\bibfield{author}{\bibinfo{person}{Edward~J Hu}, \bibinfo{person}{Yelong Shen}, \bibinfo{person}{Phillip Wallis}, \bibinfo{person}{Zeyuan Allen-Zhu}, \bibinfo{person}{Yuanzhi Li}, \bibinfo{person}{Shean Wang}, \bibinfo{person}{Lu Wang}, {and} \bibinfo{person}{Weizhu Chen}.} \bibinfo{year}{2021}\natexlab{}.
\newblock \showarticletitle{Lora: Low-rank adaptation of large language models}.
\newblock \bibinfo{journal}{\emph{arXiv preprint arXiv:2106.09685}} (\bibinfo{year}{2021}).
\newblock


\bibitem[Jiang et~al\mbox{.}(2023a)]%
        {jiang2023llmlingua}
\bibfield{author}{\bibinfo{person}{Huiqiang Jiang}, \bibinfo{person}{Qianhui Wu}, \bibinfo{person}{Chin-Yew Lin}, \bibinfo{person}{Yuqing Yang}, {and} \bibinfo{person}{Lili Qiu}.} \bibinfo{year}{2023}\natexlab{a}.
\newblock \showarticletitle{LLMLingua: Compressing Prompts for Accelerated Inference of Large Language Models}. In \bibinfo{booktitle}{\emph{Proceedings of the 2023 Conference on Empirical Methods in Natural Language Processing}}. \bibinfo{pages}{13358--13376}.
\newblock


\bibitem[Jiang et~al\mbox{.}(2023b)]%
        {jiang2023longllmlingua}
\bibfield{author}{\bibinfo{person}{Huiqiang Jiang}, \bibinfo{person}{Qianhui Wu}, \bibinfo{person}{Xufang Luo}, \bibinfo{person}{Dongsheng Li}, \bibinfo{person}{Chin-Yew Lin}, \bibinfo{person}{Yuqing Yang}, {and} \bibinfo{person}{Lili Qiu}.} \bibinfo{year}{2023}\natexlab{b}.
\newblock \showarticletitle{Longllmlingua: Accelerating and enhancing llms in long context scenarios via prompt compression}.
\newblock \bibinfo{journal}{\emph{arXiv preprint arXiv:2310.06839}} (\bibinfo{year}{2023}).
\newblock


\bibitem[Jin et~al\mbox{.}(2024)]%
        {jin2024flashrag}
\bibfield{author}{\bibinfo{person}{Jiajie Jin}, \bibinfo{person}{Yutao Zhu}, \bibinfo{person}{Xinyu Yang}, \bibinfo{person}{Chenghao Zhang}, {and} \bibinfo{person}{Zhicheng Dou}.} \bibinfo{year}{2024}\natexlab{}.
\newblock \showarticletitle{FlashRAG: A Modular Toolkit for Efficient Retrieval-Augmented Generation Research}.
\newblock \bibinfo{journal}{\emph{arXiv preprint arXiv:2405.13576}} (\bibinfo{year}{2024}).
\newblock


\bibitem[Joshi et~al\mbox{.}(2017)]%
        {joshi2017triviaqa}
\bibfield{author}{\bibinfo{person}{Mandar Joshi}, \bibinfo{person}{Eunsol Choi}, \bibinfo{person}{Daniel~S Weld}, {and} \bibinfo{person}{Luke Zettlemoyer}.} \bibinfo{year}{2017}\natexlab{}.
\newblock \showarticletitle{TriviaQA: A Large Scale Distantly Supervised Challenge Dataset for Reading Comprehension}. In \bibinfo{booktitle}{\emph{Proceedings of the 55th Annual Meeting of the Association for Computational Linguistics (Volume 1: Long Papers)}}. \bibinfo{pages}{1601--1611}.
\newblock


\bibitem[Kim et~al\mbox{.}({[n.\,d.]})]%
        {kimsure}
\bibfield{author}{\bibinfo{person}{Jaehyung Kim}, \bibinfo{person}{Jaehyun Nam}, \bibinfo{person}{Sangwoo Mo}, \bibinfo{person}{Jongjin Park}, \bibinfo{person}{Sang-Woo Lee}, \bibinfo{person}{Minjoon Seo}, \bibinfo{person}{Jung-Woo Ha}, {and} \bibinfo{person}{Jinwoo Shin}.} \bibinfo{year}{[n.\,d.]}\natexlab{}.
\newblock \showarticletitle{SuRe: Summarizing Retrievals using Answer Candidates for Open-domain QA of LLMs}. In \bibinfo{booktitle}{\emph{The Twelfth International Conference on Learning Representations}}.
\newblock


\bibitem[Kingma and Ba(2014)]%
        {kingma2014adam}
\bibfield{author}{\bibinfo{person}{Diederik~P Kingma} {and} \bibinfo{person}{Jimmy Ba}.} \bibinfo{year}{2014}\natexlab{}.
\newblock \showarticletitle{Adam: A method for stochastic optimization}.
\newblock \bibinfo{journal}{\emph{arXiv preprint arXiv:1412.6980}} (\bibinfo{year}{2014}).
\newblock


\bibitem[Kwiatkowski et~al\mbox{.}(2019)]%
        {kwiatkowski2019natural}
\bibfield{author}{\bibinfo{person}{Tom Kwiatkowski}, \bibinfo{person}{Jennimaria Palomaki}, \bibinfo{person}{Olivia Redfield}, \bibinfo{person}{Michael Collins}, \bibinfo{person}{Ankur Parikh}, \bibinfo{person}{Chris Alberti}, \bibinfo{person}{Danielle Epstein}, \bibinfo{person}{Illia Polosukhin}, \bibinfo{person}{Jacob Devlin}, \bibinfo{person}{Kenton Lee}, {et~al\mbox{.}}} \bibinfo{year}{2019}\natexlab{}.
\newblock \showarticletitle{Natural questions: a benchmark for question answering research}.
\newblock \bibinfo{journal}{\emph{Transactions of the Association for Computational Linguistics}}  \bibinfo{volume}{7} (\bibinfo{year}{2019}), \bibinfo{pages}{453--466}.
\newblock


\bibitem[Lewis et~al\mbox{.}(2020)]%
        {lewis2020retrieval}
\bibfield{author}{\bibinfo{person}{Patrick Lewis}, \bibinfo{person}{Ethan Perez}, \bibinfo{person}{Aleksandra Piktus}, \bibinfo{person}{Fabio Petroni}, \bibinfo{person}{Vladimir Karpukhin}, \bibinfo{person}{Naman Goyal}, \bibinfo{person}{Heinrich K{\"u}ttler}, \bibinfo{person}{Mike Lewis}, \bibinfo{person}{Wen-tau Yih}, \bibinfo{person}{Tim Rockt{\"a}schel}, {et~al\mbox{.}}} \bibinfo{year}{2020}\natexlab{}.
\newblock \showarticletitle{Retrieval-augmented generation for knowledge-intensive nlp tasks}.
\newblock \bibinfo{journal}{\emph{Advances in Neural Information Processing Systems}}  \bibinfo{volume}{33} (\bibinfo{year}{2020}), \bibinfo{pages}{9459--9474}.
\newblock


\bibitem[Li et~al\mbox{.}(2023a)]%
        {li2023compressing}
\bibfield{author}{\bibinfo{person}{Yucheng Li}, \bibinfo{person}{Bo Dong}, \bibinfo{person}{Frank Guerin}, {and} \bibinfo{person}{Chenghua Lin}.} \bibinfo{year}{2023}\natexlab{a}.
\newblock \showarticletitle{Compressing Context to Enhance Inference Efficiency of Large Language Models}. In \bibinfo{booktitle}{\emph{Proceedings of the 2023 Conference on Empirical Methods in Natural Language Processing}}. \bibinfo{pages}{6342--6353}.
\newblock


\bibitem[Li et~al\mbox{.}(2023b)]%
        {li2023effectiveness}
\bibfield{author}{\bibinfo{person}{Yinghui Li}, \bibinfo{person}{Haojing Huang}, \bibinfo{person}{Shirong Ma}, \bibinfo{person}{Yong Jiang}, \bibinfo{person}{Yangning Li}, \bibinfo{person}{Feng Zhou}, \bibinfo{person}{Hai-Tao Zheng}, {and} \bibinfo{person}{Qingyu Zhou}.} \bibinfo{year}{2023}\natexlab{b}.
\newblock \showarticletitle{On the (in) effectiveness of large language models for chinese text correction}.
\newblock \bibinfo{journal}{\emph{arXiv preprint arXiv:2307.09007}} (\bibinfo{year}{2023}).
\newblock


\bibitem[Liu et~al\mbox{.}(2023)]%
        {liu2023recall}
\bibfield{author}{\bibinfo{person}{Yi Liu}, \bibinfo{person}{Lianzhe Huang}, \bibinfo{person}{Shicheng Li}, \bibinfo{person}{Sishuo Chen}, \bibinfo{person}{Hao Zhou}, \bibinfo{person}{Fandong Meng}, \bibinfo{person}{Jie Zhou}, {and} \bibinfo{person}{Xu Sun}.} \bibinfo{year}{2023}\natexlab{}.
\newblock \showarticletitle{Recall: A benchmark for llms robustness against external counterfactual knowledge}.
\newblock \bibinfo{journal}{\emph{arXiv preprint arXiv:2311.08147}} (\bibinfo{year}{2023}).
\newblock


\bibitem[Ma(2019)]%
        {ma2019nlpaug}
\bibfield{author}{\bibinfo{person}{Edward Ma}.} \bibinfo{year}{2019}\natexlab{}.
\newblock \bibinfo{title}{NLP Augmentation}.
\newblock \bibinfo{howpublished}{https://github.com/makcedward/nlpaug}.
\newblock


\bibitem[Ma et~al\mbox{.}(2023)]%
        {ma2023query}
\bibfield{author}{\bibinfo{person}{Xinbei Ma}, \bibinfo{person}{Yeyun Gong}, \bibinfo{person}{Pengcheng He}, \bibinfo{person}{Hai Zhao}, {and} \bibinfo{person}{Nan Duan}.} \bibinfo{year}{2023}\natexlab{}.
\newblock \showarticletitle{Query Rewriting in Retrieval-Augmented Large Language Models}. In \bibinfo{booktitle}{\emph{Proceedings of the 2023 Conference on Empirical Methods in Natural Language Processing}}. \bibinfo{pages}{5303--5315}.
\newblock


\bibitem[Mallen et~al\mbox{.}(2022)]%
        {mallen2022not}
\bibfield{author}{\bibinfo{person}{Alex Mallen}, \bibinfo{person}{Akari Asai}, \bibinfo{person}{Victor Zhong}, \bibinfo{person}{Rajarshi Das}, \bibinfo{person}{Hannaneh Hajishirzi}, {and} \bibinfo{person}{Daniel Khashabi}.} \bibinfo{year}{2022}\natexlab{}.
\newblock \showarticletitle{When not to trust language models: Investigating effectiveness and limitations of parametric and non-parametric memories}.
\newblock \bibinfo{journal}{\emph{arXiv preprint arXiv:2212.10511}}  \bibinfo{volume}{7} (\bibinfo{year}{2022}).
\newblock


\bibitem[Muennighoff et~al\mbox{.}(2022)]%
        {muennighoff2022mteb}
\bibfield{author}{\bibinfo{person}{Niklas Muennighoff}, \bibinfo{person}{Nouamane Tazi}, \bibinfo{person}{Lo{\"\i}c Magne}, {and} \bibinfo{person}{Nils Reimers}.} \bibinfo{year}{2022}\natexlab{}.
\newblock \showarticletitle{MTEB: Massive Text Embedding Benchmark}.
\newblock \bibinfo{journal}{\emph{arXiv preprint arXiv:2210.07316}} (\bibinfo{year}{2022}).
\newblock
\urldef\tempurl%
\url{https://doi.org/10.48550/ARXIV.2210.07316}
\showDOI{\tempurl}


\bibitem[Saad-Falcon et~al\mbox{.}(2024)]%
        {saad2024ares}
\bibfield{author}{\bibinfo{person}{Jon Saad-Falcon}, \bibinfo{person}{Omar Khattab}, \bibinfo{person}{Christopher Potts}, {and} \bibinfo{person}{Matei Zaharia}.} \bibinfo{year}{2024}\natexlab{}.
\newblock \showarticletitle{ARES: An Automated Evaluation Framework for Retrieval-Augmented Generation Systems}. In \bibinfo{booktitle}{\emph{Proceedings of the 2024 Conference of the North American Chapter of the Association for Computational Linguistics: Human Language Technologies (Volume 1: Long Papers)}}. \bibinfo{pages}{338--354}.
\newblock


\bibitem[Shao et~al\mbox{.}(2023)]%
        {shao2023enhancing}
\bibfield{author}{\bibinfo{person}{Zhihong Shao}, \bibinfo{person}{Yeyun Gong}, \bibinfo{person}{Yelong Shen}, \bibinfo{person}{Minlie Huang}, \bibinfo{person}{Nan Duan}, {and} \bibinfo{person}{Weizhu Chen}.} \bibinfo{year}{2023}\natexlab{}.
\newblock \showarticletitle{Enhancing Retrieval-Augmented Large Language Models with Iterative Retrieval-Generation Synergy}. In \bibinfo{booktitle}{\emph{Findings of the Association for Computational Linguistics: EMNLP 2023}}. \bibinfo{pages}{9248--9274}.
\newblock


\bibitem[Shi et~al\mbox{.}(2024)]%
        {shi2024replug}
\bibfield{author}{\bibinfo{person}{Weijia Shi}, \bibinfo{person}{Sewon Min}, \bibinfo{person}{Michihiro Yasunaga}, \bibinfo{person}{Minjoon Seo}, \bibinfo{person}{Richard James}, \bibinfo{person}{Mike Lewis}, \bibinfo{person}{Luke Zettlemoyer}, {and} \bibinfo{person}{Wen-tau Yih}.} \bibinfo{year}{2024}\natexlab{}.
\newblock \showarticletitle{REPLUG: Retrieval-Augmented Black-Box Language Models}. In \bibinfo{booktitle}{\emph{Proceedings of the 2024 Conference of the North American Chapter of the Association for Computational Linguistics: Human Language Technologies (Volume 1: Long Papers)}}. \bibinfo{pages}{8364--8377}.
\newblock


\bibitem[Tonmoy et~al\mbox{.}(2024)]%
        {tonmoy2024comprehensive}
\bibfield{author}{\bibinfo{person}{SM Tonmoy}, \bibinfo{person}{SM Zaman}, \bibinfo{person}{Vinija Jain}, \bibinfo{person}{Anku Rani}, \bibinfo{person}{Vipula Rawte}, \bibinfo{person}{Aman Chadha}, {and} \bibinfo{person}{Amitava Das}.} \bibinfo{year}{2024}\natexlab{}.
\newblock \showarticletitle{A comprehensive survey of hallucination mitigation techniques in large language models}.
\newblock \bibinfo{journal}{\emph{arXiv preprint arXiv:2401.01313}} (\bibinfo{year}{2024}).
\newblock


\bibitem[Wang et~al\mbox{.}(2023)]%
        {wang2023query2doc}
\bibfield{author}{\bibinfo{person}{Liang Wang}, \bibinfo{person}{Nan Yang}, {and} \bibinfo{person}{Furu Wei}.} \bibinfo{year}{2023}\natexlab{}.
\newblock \showarticletitle{Query2doc: Query Expansion with Large Language Models}. In \bibinfo{booktitle}{\emph{Proceedings of the 2023 Conference on Empirical Methods in Natural Language Processing}}. \bibinfo{pages}{9414--9423}.
\newblock


\bibitem[Wolf et~al\mbox{.}(2020)]%
        {wolf2020transformers}
\bibfield{author}{\bibinfo{person}{Thomas Wolf}, \bibinfo{person}{Lysandre Debut}, \bibinfo{person}{Victor Sanh}, \bibinfo{person}{Julien Chaumond}, \bibinfo{person}{Clement Delangue}, \bibinfo{person}{Anthony Moi}, \bibinfo{person}{Pierric Cistac}, \bibinfo{person}{Tim Rault}, \bibinfo{person}{R{\'e}mi Louf}, \bibinfo{person}{Morgan Funtowicz}, {et~al\mbox{.}}} \bibinfo{year}{2020}\natexlab{}.
\newblock \showarticletitle{Transformers: State-of-the-art natural language processing}. In \bibinfo{booktitle}{\emph{Proceedings of the 2020 conference on empirical methods in natural language processing: system demonstrations}}. \bibinfo{pages}{38--45}.
\newblock


\bibitem[Xiao et~al\mbox{.}(2023)]%
        {xiao2023c}
\bibfield{author}{\bibinfo{person}{Shitao Xiao}, \bibinfo{person}{Zheng Liu}, \bibinfo{person}{Peitian Zhang}, \bibinfo{person}{Niklas Muennighoff}, \bibinfo{person}{Defu Lian}, {and} \bibinfo{person}{Jian-Yun Nie}.} \bibinfo{year}{2023}\natexlab{}.
\newblock \showarticletitle{C-pack: Packaged resources to advance general chinese embedding}.
\newblock \bibinfo{journal}{\emph{arXiv preprint arXiv:2309.07597}} (\bibinfo{year}{2023}).
\newblock


\bibitem[Xu et~al\mbox{.}(2024)]%
        {xu2024sparsecl}
\bibfield{author}{\bibinfo{person}{Haike Xu}, \bibinfo{person}{Zongyu Lin}, \bibinfo{person}{Yizhou Sun}, \bibinfo{person}{Kai-Wei Chang}, {and} \bibinfo{person}{Piotr Indyk}.} \bibinfo{year}{2024}\natexlab{}.
\newblock \showarticletitle{SparseCL: Sparse Contrastive Learning for Contradiction Retrieval}.
\newblock \bibinfo{journal}{\emph{arXiv preprint arXiv:2406.10746}} (\bibinfo{year}{2024}).
\newblock


\bibitem[Yang et~al\mbox{.}(2024)]%
        {yang2024qwen2}
\bibfield{author}{\bibinfo{person}{An Yang}, \bibinfo{person}{Baosong Yang}, \bibinfo{person}{Binyuan Hui}, \bibinfo{person}{Bo Zheng}, \bibinfo{person}{Bowen Yu}, \bibinfo{person}{Chang Zhou}, \bibinfo{person}{Chengpeng Li}, \bibinfo{person}{Chengyuan Li}, \bibinfo{person}{Dayiheng Liu}, \bibinfo{person}{Fei Huang}, {et~al\mbox{.}}} \bibinfo{year}{2024}\natexlab{}.
\newblock \showarticletitle{Qwen2 technical report}.
\newblock \bibinfo{journal}{\emph{arXiv preprint arXiv:2407.10671}} (\bibinfo{year}{2024}).
\newblock


\bibitem[Yang et~al\mbox{.}(2018)]%
        {yang2018hotpotqa}
\bibfield{author}{\bibinfo{person}{Zhilin Yang}, \bibinfo{person}{Peng Qi}, \bibinfo{person}{Saizheng Zhang}, \bibinfo{person}{Yoshua Bengio}, \bibinfo{person}{William Cohen}, \bibinfo{person}{Ruslan Salakhutdinov}, {and} \bibinfo{person}{Christopher~D Manning}.} \bibinfo{year}{2018}\natexlab{}.
\newblock \showarticletitle{HotpotQA: A Dataset for Diverse, Explainable Multi-hop Question Answering}. In \bibinfo{booktitle}{\emph{Proceedings of the 2018 Conference on Empirical Methods in Natural Language Processing}}. Association for Computational Linguistics.
\newblock


\bibitem[Ye et~al\mbox{.}(2023)]%
        {ye2023improving}
\bibfield{author}{\bibinfo{person}{Dezhi Ye}, \bibinfo{person}{Bowen Tian}, \bibinfo{person}{Jiabin Fan}, \bibinfo{person}{Jie Liu}, \bibinfo{person}{Tianhua Zhou}, \bibinfo{person}{Xiang Chen}, \bibinfo{person}{Mingming Li}, {and} \bibinfo{person}{Jin Ma}.} \bibinfo{year}{2023}\natexlab{}.
\newblock \showarticletitle{Improving Query Correction Using Pre-train Language Model In Search Engines}. In \bibinfo{booktitle}{\emph{Proceedings of the 32nd ACM International Conference on Information and Knowledge Management}}. \bibinfo{pages}{2999--3008}.
\newblock


\bibitem[Zhang et~al\mbox{.}(2024b)]%
        {zhang2024trigger}
\bibfield{author}{\bibinfo{person}{Kepu Zhang}, \bibinfo{person}{Zhongxiang Sun}, \bibinfo{person}{Xiao Zhang}, \bibinfo{person}{Xiaoxue Zang}, \bibinfo{person}{Kai Zheng}, \bibinfo{person}{Yang Song}, {and} \bibinfo{person}{Jun Xu}.} \bibinfo{year}{2024}\natexlab{b}.
\newblock \showarticletitle{Trigger$^3$: Refining Query Correction via Adaptive Model Selector}.
\newblock \bibinfo{journal}{\emph{arXiv preprint arXiv:2412.12701}} (\bibinfo{year}{2024}).
\newblock


\bibitem[Zhang et~al\mbox{.}(2024a)]%
        {zhang2024raft}
\bibfield{author}{\bibinfo{person}{Tianjun Zhang}, \bibinfo{person}{Shishir~G Patil}, \bibinfo{person}{Naman Jain}, \bibinfo{person}{Sheng Shen}, \bibinfo{person}{Matei Zaharia}, \bibinfo{person}{Ion Stoica}, {and} \bibinfo{person}{Joseph~E Gonzalez}.} \bibinfo{year}{2024}\natexlab{a}.
\newblock \showarticletitle{Raft: Adapting language model to domain specific rag}.
\newblock \bibinfo{journal}{\emph{arXiv preprint arXiv:2403.10131}} (\bibinfo{year}{2024}).
\newblock


\end{thebibliography}

\appendix

\end{document}